\documentclass[aps,prd,showpacs,nofootinbib,twocolumn,superscriptaddress,floatfix]{revtex4}

\usepackage{amsmath,amssymb}
\usepackage[dvips, pdftex]{graphicx}
\usepackage[tight,footnotesize]{subfigure}
\usepackage[usenames]{color}
\usepackage[dvipsnames]{xcolor}
\usepackage{float}
\usepackage{wrapfig}
\usepackage{hyperref}
\usepackage{mathtools}

\usepackage[nameinlink]{cleveref}

\DeclareGraphicsExtensions{.jpg,.mps,.pdf,.png,.eps}

\DeclareMathOperator{\arcsinh}{arcsinh}
\DeclareMathOperator{\arctanh}{arctanh}

\renewcommand{\ij}{{}_{ij}} 
\renewcommand{\IJ}{{}^{ij}}

\newcommand{\gm}{\gamma}

\newcommand{\lp}{\left(}
\newcommand{\rp}{\right)}

\newcommand{\tgm}{{\tilde\gm}}

\newcommand{\tA}{{\tilde A}}

\newcommand{\be}{\begin{equation}}
\newcommand{\ee}{\end{equation}}
\newcommand{\bea}{\begin{eqnarray}}
\newcommand{\eea}{\end{eqnarray}}

\newcommand{\beq}{\begin{equation}}
\newcommand{\eeq}{\end{equation}}

\newcommand{\rr}{\mathrm}

\newcommand{\Mpl}{M_\rr{pl}}

\begin{document}
\title{
Gravitational dynamics in Higgs inflation: \\ Preinflation and preheating with an auxiliary field
}

\author{Cristian Joana}
\email{cristian.joana@uclouvain.be}
\affiliation{Cosmology, Universe and Relativity at Louvain (CURL),
	Institute of Mathematics and Physics,
	University of Louvain,
	Chemin du Cyclotron 2,
	1348 Louvain-la-Neuve,
	Belgium}
\affiliation{Service de Physique Th\'eorique, Universit\'e Libre de Bruxelles (ULB), Boulevard du Triomphe, CP225, 1050 Brussels, Belgium.}

\pacs{98.80.Cq, 98.70.Vc}
	
\date{\today}

\begin{abstract}
The dynamics of both the preinflationary and the preheating epochs for a model consisting of a Higgs inflaton plus an additional auxiliary field are studied in full General Relativity. The minimally coupled auxiliary field allows for parametric-type resonances that successfully transfer energy from the inflaton condensate to particle excitations in both fields. 
Depending on the interaction strengths of the fields, the broad resonance periods lead to structure formation consisting of large under/over-densities, and possibly the formation of compact objects. Moreover, when confronting the same model to multi-field inhomogeneous preinflation, the onset of inflation is shown to be a robust outcome. 
At relatively large Higgs values, the non-minimal coupling acts as a stabilizer, protecting the dynamics of the inflaton, and significantly reducing the impact of perturbations in other fields and matter sectors.  These investigations further confirm the robustness of Higgs inflation to multi-field inhomogeneous initial conditions, while putting in evidence the formation of complex structures during the reheating.

\end{abstract}

\maketitle

\section{Introduction}\label{sec:Intro}

Cosmic inflation \cite{STAROBINSKY198099, PhysRevD.23.347, 10.1093/mnras/195.3.467, LINDE1982389} is the current paradigm of the early Universe. It postulates an early phase where the Universe underwent over a large period of accelerated expansion. Such a period provides an explanation for today's large scale homogeneity and flatness of the Universe.
During inflation, quantum fluctuations became red-shifted, exiting the Hubble horizon at the time, and leading to a scale-invariant power spectrum of cosmological perturbations which can be matched to current observations \cite{Akrami:2018odb,Ade:2015lrj}. At later times, they provide the seeds needed for structure formation. 

In a universe governed by the Einstein's field equations, the accelerated expansion of the universe is obtained when the effective equation of state is strictly smaller than $\omega < -1/3$. In the slow-roll inflationary paradigm, this is typically achieved by postulating a universe dominated by a scalar field (slowly) rolling down its potential.  
%
Assuming homogeneity and isotropy, from the shape of the potential, the slow-roll conditions can be derived. When these conditions are satisfied, the energy budget is dominated by the potential energy (i.e it keeps an $\omega \approx -1$), and a sustained period of slow-roll inflation occurs. Surely, assuming homogeneity and isotropy for the initial conditions of the universe is one of the main problem inflation is supposed to solve, so the initial conditions required for inflation have often been a topic of controversy, e.g. in Refs.~\cite{Goldwirth:1989pr, Goldwirth:1990pm, Laguna:1991zs,KurkiSuonio:1993fg,Deruelle:1994pa} and more recently in Refs.~\cite{Martin:2013nzq, Ijjas_2013,Guth_2014,Easther:2014zga, Ijjas_2016,Chowdhury:2019otk}. Thus, the remaining question is; Can generic (inhomogeneous) preinflationary scenarios successfully lead to enough cosmic inflation ($\sim 60$ \text{efolds})?  

The issue of initial conditions for inflation has been studied extensively using analytical, semi-analytical and numerical approaches (for a review see Ref.~\cite{Brandenberger:2016uzh}). Full numerical relativity simulations have also been used to explore the dynamics of the preinflationary era beyond the perturbative regime. These have consisted of scenarios with a highly inhomogeneous scalar field \cite{East:2015ggf,Clough:2017ixw} and large tensor perturbations \cite{Clough_2018}.  The effects of concave and convex potential shapes were also studied in Ref.~\cite{Aurrekoetxea_2020}.

In our previous paper \cite{Joana2020}, the case of (single field) Higgs/Starobinsky preinflation was considered, containing large field gradients and inhomogeneous kinetic energies across Hubble scales. We have shown that for this model, gravitational shear and tensor modes can potentially delay the onset of inflation, but never prevent it. The question of the implications of adding extra fields is, however, still open. 

The (p)reheating epoch is a necessary phase occurring after the end of inflation. It starts once the slow-roll conditions are violated and the inflaton condensate begins to oscillate around the minimum of its potential. These oscillations transfer energy to the matter sector, through parametric resonances, originating in the hot big bang plasma \cite{PhysRevD.42.2491,1990Dolgov172D}. Usually, in the literature, the phase when particles are produced is known as ``preheating'', while the term ``reheating'' is left for when the inflaton has effectively decayed and the thermalization phase begins. 
The reheating process has direct implications on the Cosmic Microwave Background (CMB), and current measurements are sensitive to it \cite{Martin2010,Martin2015,Martin2016}. 
For a more elaborate review on the topic, see Refs.~\cite{Brandenberger2010, Tenkanen:2020cvw}.  

The dynamics of the initial stages of preheating have been extensively studied throughout the last decade. Perturbative approaches \cite{PhysRevD.42.2491,1990Dolgov172D,Kofman1994,Kofman1997, Tsujikawa1999A, Tsujikawa1999B} and numerical lattice simulations \cite{Tomislav1997,Gary2001,Bellido2003,Amin2010, Frolov2010,Lozanov2014,Repond2016,Lozanov2019} have been used extensively while assuming linearized Einstein gravity.  Reheating involving non-minimally coupled scalar fields has also been of large interest \cite{Tsujikawa1999A, Tsujikawa1999B, Bruck2017, DeCross2018A,DeCross2018B, DeCross2018C,Sfakianakis2019,Nguyen2019,Rubio2019,Vis2020,Ema2017,Ema2021}, and includes studies of Higgs inflation \cite{Repond2016,Sfakianakis2019,Rubio2019,Hamada2021}. While lattice simulations have been capable to preserve the non-perturbative dynamics associated with inhomogeneous scalar fields, they do not consider the fully non-linear gravitational counterparts \cite{Bassett1999}, whose effects on the structure formation might lead to the early formation of black holes \cite{Jedamzik2010A,Jedamzik2010B,Zihan2020}. In 2019, Giblin and Tishue, in Ref.~\cite{Giblin2019} presented the first preheating simulations in full general relativity for the canonical $m^2\varphi^2$ inflationary model. 
While their results disfavor the formation of compact structures, for that particular model, it shows the potential of numerical relativity to clarify the role of gravitational backreactions in the early Universe, complementary to the standard cosmological perturbation theory. In 2020, Kou \textit{et. al.} in Ref.~\cite{Kou:2019bbc} (see also \cite{Kou:2021bij}) presented numerical relativity simulations for an alternative inflationary model which allowed the formation of oscillons during the preheating potentially collapsing into black holes. 

In this paper, I present a set of full general relativity simulations concerning both the preinflationary and preheating epochs. The non-minimally coupled Higgs inflation model has been considered in the presence of an auxiliary scalar field. 
With the help of these simulations, we first ask ourselves how a full general relativistic treatment affects the resonant dynamics of preheating, and how the coupling strength of the fields affect the formation of structures during the broad resonance phase. Then, we check whether similar dynamics can be present during the preinflationary phase and, importantly, if these can undermine  the beginning of inflation in the first place.  It is shown that, in the presence of additional fields, the non-minimal coupling  to gravity of the Higgs field allows for an efficient preheating process; a large amount of particles are produced and the formation of complex structures occurs. However, during preinflation, at large enough Higgs field values,  the non-minimal coupling always acts as a stabilizer that protects the dynamics of the inflaton from inhomogeneities in other fields, ensuring the success of starting cosmic inflation. 

The organization of the manuscript is as follows: in Section \ref{sec:Formalism} the generalized covariant formalism is introduced while in Section \ref{sec:HiggsModel}  focus on the Higgs model. %
Section \ref{sec:NumStrategy} explains the numerical strategy of the simulations. The results for preheating and preinflation are presented in Sections \ref{sec:SimsReheat} and \ref{sec:SimsPreinf}, respectively. Additional information on the notation, code performance, initial data sets and supplementary figures are available in the appendixes.

\section{Covariant formalism}\label{sec:Formalism}

In this section, we consider a universe containing an arbitrary number of scalar fields $\bar \phi^I$, labeled by Latin capital letters $I, J, K = 1, 2, ... , N$. We consider a metric tensor $\bar g^{\mu\nu}$ in $3+1$ dimensions where Greek letters are used to label spacetime indices $\mu, \nu = 0, 1, 2, 3$, using the ``mostly plus metric'' sign convention $(-+++)$. The variables with an upper-bar or ``hat'' are being described in the Jordan frame. In these kind of models, the action in the Jordan frame is given by 
\beq
\begin{split}
S = \int d^4 x & \sqrt{-\bar{g} } \Big[ f (\bar\phi^I ) \bar{R} 
\\
& - \frac{\Mpl^2}{2} \delta_{IJ} \bar{g}^{\mu\nu} \partial_\mu \bar\phi^I \partial_\nu \bar\phi^J - U(\bar\phi^I ) \Big] ~ ,
\end{split}
\label{eq:action_J}
\eeq
where $\Mpl$ is the reduced Planck mass, $\bar g$ is the determinant of the metric, $\bar R$ is the Ricci scalar, $U(\bar\phi^I)$ is the scalar field potential, and $f(\bar\phi^I)$ contains the fields non-minimal coupling gravity $\xi_I$, so that

\beq \label{eq:f_func}
 f (\bar\phi^I ) = \frac {\Mpl^2} 2 \left[ 1 + \xi_K \left(\bar\phi^K\right)^2 \right] ~.
\eeq

The dynamical analysis of such systems is easier to deal with in the Einstein frame. This is done by rescaling the metric tensor, under the Weyl transformation 
\bea
\bar{g}_{\mu\nu} (x) \rightarrow g_{\mu\nu} (x) = \frac{2}{M_{\rm pl}^2} f \big(\bar\phi^I\big) \> \bar{g}_{\mu \nu} (x) ~.
\eea

Thus, now in the Einstein frame, the action reads 
\beq 
\begin{split}
S = \int d^4 x \sqrt{-g} \Big[ R & - \frac{\Mpl^2}{2} {\cal G}_{IJ} (\bar\phi^K ) g^{\mu\nu} \partial_\mu \bar\phi^I \partial_\nu \bar\phi^J
\\
& - V (\bar\phi^I ) \Big] \, ,
\end{split}
\label{eq:action_E}
\eeq 

Where ${\cal G}_{IJ} (\phi^K)$ is a field-space metric containing the mixing with the non-minimal coupling, 
\beq
{\cal G}_{IJ} (\bar\phi^K ) = \frac{ M_{\rm pl}^2 }{2 f (\bar\phi^K) }\left[ \delta_{IJ} + \frac{3}{ f (\bar\phi^K) } \frac{\partial f} {\partial \bar\phi^I} \frac{\partial f} {\partial \bar\phi^J} \right] ~,
\label{eq:G_IJ}
\eeq

and the field potential has been redefined as 
\beq
V (\bar\phi^I) = \frac{\Mpl^4}{ 4 f^2 (\bar\phi^I) } U(\bar\phi^I ) .
\label{VE}
\eeq

Varying the action of Eq.~(\ref{eq:action_E}) with respect to $\phi^I$, one can find the stress tensor and the field's equations of motion:
\beq
T_{\mu\nu} = {\cal G}_{IJ} \partial_\mu \bar\phi^I \partial_\nu \bar\phi^J - g_{\mu\nu} \left[ \frac{1}{2} {\cal G}_{IJ} \partial_\alpha \bar\phi^I \partial^\alpha \bar\phi^J + V (\bar\phi^I ) \right] ~,
\label{eq:Tmn_J}
\eeq
\beq
\Box \bar\phi^I + g^{\mu\nu} \Gamma^I_{JK} \partial_\mu \bar\phi^J \partial_\nu \bar\phi^K - {\cal G}^{IJ} \frac{\partial }{\partial {\bar\phi}^J} V (\bar\phi^K) = 0 ,
\label{eq:eomsf_J}
\eeq

where $\Box$ is the Alembertian operator, and $\Gamma^I_{JK} (\phi^L)$ are the Christoffel symbols constructed from the field-space metric ${\cal G}_{IJ}$.

The canonical Einstein fields denoted by $\Phi^I$ are defined by solving the following system of equations 
\beq \label{eq:convert_frame}
\frac{\Mpl^2}{2} {\cal G}_{IJ} g^{\mu\nu} \partial_\mu \bar\phi^I \partial_\nu \bar\phi^J = {\delta}_{IJ} g^{\mu\nu} \partial_\mu \Phi^I \partial_\nu \Phi^J ~.
\eeq
This transformation further simplifies the action in Eq.~(\ref{eq:action_E}). However, finding the solution to such a system of equations is not always straightforward
\footnote{
In fact, finding a global transformation that solves Eqs. \ref{eq:convert_frame} is not possible when the field-space manifold  ${\cal G}^{IJ}$ is curved \cite{DeCross2018A}. However, it is often possible to find an approximate local solution that maps both field spaces 
in a specific region of the field space 
(see section \ref{sec:conversion_frames}).}.

Thereafter, in the Einstein frame, the field equations of motion are reduced to the classical Klein-Gordon equations of the form 
\beq
\Box \Phi^I -\frac{\partial }{\partial \Phi^I} V (\Phi^K) = 0 ~.
\eeq

\section{Higgs Inflation}\label{sec:HiggsModel}

In this work we consider the model of (non-minimally coupled) Higgs inflation which
is one of the most favored slow-roll inflation models by the latest CMB data from Planck~\cite{Martin:2013nzq}. We consider a dynamical system consisting of two scalar fields and gravity. Interaction between the Higgs field and other Standard Model particles, particularly in the electroweak sector, have been ignored. The evolution has been treated classically, therefore radiative loop corrections have also been neglected. Section \ref{subsec:SingleHiggs}, 
briefly reviews the formalism for the single-field paradigm, assuming the unitary gauge, while  the implications of adding extra scalar fields are discussed in Sec.~\ref{subsec:AuxField}.

\subsection{The single-field case } \label{subsec:SingleHiggs}

The Higgs inflation model~\cite{Bezrukov:2007ep} postulates that the inflaton is the Higgs field from the Standard Model of particle physics, with a non-minimal coupling to gravity. The Standard Model Lagrangian, therefore, includes an extra term $\xi H^\dagger H R$, where $R$ is the Ricci scalar, and $H$ is the Higgs field in the unitary gauge \cite{Bellido2009}, 
\beq
H = \frac{\Mpl}{\sqrt{2}}
\begin{pmatrix} 
0 \\ h
\end{pmatrix} 
\eeq
and $\xi_h$ is the only free  parameter of the model. This term is somehow expected as it is  naturally generated by quantum corrections in curved spacetime \cite{1970Callan}. 
\\

\begin{figure}[!t]
\begin{center}
\hspace*{-5mm}
\includegraphics[width=9.5cm]{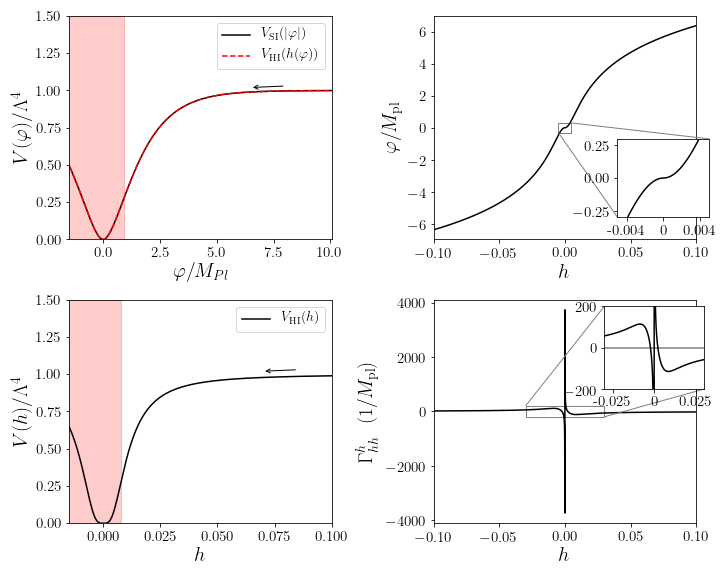}
\caption{Left panels illustrate the Higgs potential in the Einstein frame in terms of the $\varphi$-field (top-left) and the Higgs field $h$ (bottom-left). Slow-roll inflation runs from right to left as indicated by the arrow; the red-shaded area indicates the post-inflationary period, after the first slow-roll parameter becomes larger than unity.  The top-right panel shows the conversion between $h$ and $\varphi$, and in the bottom-right panel, the field-space Christoffel symbol $\Gamma^h_{hh}$ is plotted to illustrate the kinematic factor felt by $h$ due to the non-minimal coupling, as seen in the Einstein frame, (see Eq.~\ref{eq:eomsf_J}). The rapid field-accelerations occurring around $h\approx 0$ result into the so-called  Riemann spikes observed during the evolution.}
\label{fig:Higgs_EJ}
\end{center}
\end{figure}

In the Einstein frame, the Higgs potential reads  
\beq \label{VHI_pot}
V(h) = \Mpl^4 \frac{\lambda \left( h^2 - \frac{v^2}{\Mpl } \right)^2}{4 \left( 1 + \xi_h h^2\right)^2} ~,
\eeq
the shape of which is illustrated in Fig. \ref{fig:Higgs_EJ}. For the single field case, using Eq.~(\ref{eq:convert_frame}), one can  convert from the Jordan frame field $h$  to the canonical inflaton in the Einstein frame $\varphi$  by solving

\beq \label{eq:diff_varphi_of_h}
 \frac 1 {\Mpl} \frac{\text{d} \varphi} {\text{d} h} = \sqrt{{\cal G}_{hh}} = \frac{\sqrt{1+\xi_h(1+6\xi_h)h^2}}{1+\xi_h h^2} ~,
\eeq
which leads to the known expression \cite{MARTIN201475}

\beq
\begin{split} \label{eq:varphi_of_h}
\frac{\varphi}{ \Mpl} = &\sqrt{\frac{1+6\xi_h}{\xi_h}} \arcsinh\left({h \sqrt{\xi_h (1+6\xi_h)}}\right) \\ &- \sqrt{6} \arctanh\left({\frac{\xi_h\sqrt{6} h}{\sqrt{1+\xi_h(1+6\xi_h)h^2}}}\right) ~.
\end{split}
\eeq

Expanding the above expression and substituting it in the potential (\ref{VHI_pot}), one gets, in terms of the $\varphi$-field,  

\beq \label{eq:V_SI}
V_{\rm SI}(\varphi) \approx \Lambda^4 \left( 1 - {\rm e}^{-\sqrt{2/3} |\varphi| / \Mpl} \right)^2~,
\eeq
where
\beq
\Lambda^4 \equiv \Mpl^4 \lambda / (4 \xi_h^2) ~,
\eeq
is the overall amplitude of the potential. 
\\

The energy scale of inflation is given by the amplitude of the potential, $H^2_{\rm inf} \approx \Lambda^4/(3\Mpl)$. Assuming that  the observable modes exited the Hubble radius at $N_\star = 55$ efolds  before the end of inflation, the scalar and tensor perturbations of the CMB power-spectrum lead to  ${\Lambda\simeq 3.1 \times 10^{-3} \Mpl} $ ~\cite{Martin:2013nzq}. Thus, the ratio between the Higgs self-coupling and the non-minimal coupling must obey
\beq \label{eq:ratio_coupling}
\frac \lambda {\xi_h^2} \simeq 5 \cdot 10^{-10} ~.
\eeq
Ignoring the running the Higgs self-coupling \cite{Degrassi_2012}, this is set to the measured value by collider physics,  
$\lambda \simeq 0.13$ \cite{PhysRevD.98.030001}. 
In such a case, {{Eq.~(\ref{eq:ratio_coupling})}} fixes the value of the Higgs non-minimal coupling to ${\xi_h \approx 1.8 \cdot 10^4}$.
\\

At leading order, the firsts two slow-roll parameters read 
\begin{align}
\epsilon_1 &\simeq \frac{\Mpl^2}2 \left( \frac{\partial_h V}{V}\right)^2   ~,
\\ %
\epsilon_2 &\simeq 2\Mpl^2 \left[ \left(\frac{\partial_h V}{V}\right)^2 - \frac{\partial^2_h \, V}{V} \right] ~, 
\end{align}
and as long $\epsilon_1 <1$ (homogeneous) inflation is granted. In other words, the inflationary trajectory ends when $\epsilon = 1$, corresponding to an equation of state $\omega = -1/3 $. Assuming that inflation lasted, at least, the minimum amount to explain the CMB observations, ${\Delta N \simeq 55}$ efolds, this implies that  it should have started at a field value of  $\varphi_* \gtrsim 5.5\ \Mpl$ ($h_* \gtrsim 0.1$). Once cosmic inflation takes place, the field slowly rolls down the potential until the kinetic energy breaks the slow-roll conditions. The end of inflation occurs approximately at $\varphi_{\rm end} \approx 0.94\> \Mpl$ ($h_{\rm end} \approx 0.008$) ~\cite{MARTIN201475}, signifying the beginning of the reheating epoch.

\subsection{Higgs with an auxiliary field }\label{subsec:AuxField}

Let us consider now the addition of an auxiliary field $s$, into the Higgs inflation model. To keep within the spirit of the original single-field model, in this paper, we restrict ourselves to the case where the auxiliary field is minimally coupled to gravity ($\xi_s = 0$).  On the other hand, an interaction term is added in the action of Eq.~(\ref{eq:action_E}), 
\beq \label{eq:L_int}
{\cal L}_\text{int} = - {\mathsf{g}\,} h^2s^2 
\eeq
where $\mathsf{g}$ is the field-field coupling constant. This term is necessary for a parametric-type preheating to occur at the end of inflation. After this  modification, the potential in {Eq.~(\ref{eq:V_SI})} becomes
\beq \label{eq:V_HIaux}
V(h, s) ={\Mpl^4} \frac { \left[ \frac{\lambda}4 \left( h^2 - {v^2}/{\Mpl^2} \right)^2 + {\mathsf{g}\,} h^2 s^2 \right]} 
{ \left(1 + \xi_h h^2  \right)^2}~.
\eeq

It is  relevant to note that  the effect of the non-minimal coupling $\xi_h$ on the potential is crucial. While in the Jordan frame the potential becomes larger $U (h, s) \rightarrow \infty$ at large Higgs-values $ h \rightarrow \infty$, in the Einstein frame the potential tends to the constant plateau $ V(h,s) \rightarrow \Lambda^4$, effectively suppressing the interaction term and stabilizing the dynamics. This effect applies as well to any other possible coupling between the inflaton and other matter sources, including high energy new physics \cite{Branchina2019}, which remarkably generalizes the dynamics at large field values, and thus, during preinflation \cite{Linde:1983mx,Linde:2017pwt}. 
\\

In this extension of the model, the canonically normalized fields in the Einstein frame are denoted by $\varphi,~\chi$. Where $\varphi$ represents the inflaton, and $\chi$ the auxiliary field.  

\subsection{Conversion between the fields in the Jordan and Einstein frame notation}  \label{sec:conversion_frames}

The fact that the $s$-field is assumed to be minimally coupled, 
facilitates the analysis as it simplifies the mixing between the fields and gravity. Indeed, under this assumption, the field-space metric becomes diagonal, ${\cal G}_{IJ} = {\rm diag} ({\cal G}_{hh}, {\cal G}_{ss})$.  This is convenient because allows us to easily infer the momentum of the Einstein framed fields ($\Pi_\varphi,\ \Pi_\chi$)  the Jordan ones ($\Pi_h,\ \Pi_s$), by  (no index-summation implied) 
\beq
\Pi_\varphi^2 = {\cal G}_{hh} \Pi_h ^2 
~, \qquad 
\Pi_\chi^2 = {\cal G}_{ss} \Pi_s ^2 ~.
\eeq

In principle, the conversion of the field values should be done by solving {Eq.~(\ref{eq:convert_frame})}. 
However, at small-field values the conversion can be well approximated by solving 
\begin{align} \label{eq:approx_fieldconvers}
  \frac 1 {\Mpl} \frac{\partial \varphi} {\partial h} \approx \sqrt{{\cal G}_{hh}}  ~, \qquad 
   \frac 1 {\Mpl} \frac{\partial \chi} {\partial s}  \approx  \sqrt{{\cal G}_{ss}}  ~,
\end{align}
recovering  {Eq.~(\ref{eq:varphi_of_h})} for the inflaton, while the auxiliary  field in that frame 
is approximately given by
\beq \label{eq:phi_of_s}
\frac \chi \Mpl \approx \left( 1 + \xi_h h^2\right)^{-1/2} s ~.
\eeq

Note that, as shown in Appendix~\ref{ApConversionJE}, these approximations are not valid in some parts of the field-space when $s \gtrsim 0.1$, therefore they cannot be used when large field excursions are present, such as when considering preinflationary scenarios (see Sec.~\ref{sec:numdetails}).

\section{Numerical strategy}\label{sec:NumStrategy}

The end goal of this paper is to test if Higgs inflation in the presence of an auxiliary field can begin from inhomogeneous initial conditions, and samewise if it is able to preheat the universe after the end of inflation via parametric preheating. To that end, I will be using the GRChombo numerical relativity code ~\cite{Clough_2015,Andrade2021} to simulate the pre- and post-inflationary dynamics in full general relativity. %
\\

In the 3+1 decomposition of General Relativity ~\cite{Gourgoulhon:2007ue} the line element is written as 
\bea \label{timeline}
\rr d s^2 = - \alpha^2 \rr d t^2 + \frac1\chi \tgm\ij (\rr d x^i + \beta^i \rr d t)(\rr d x^j + \beta^j \rr d t)~,
\eea
where it has been used the conformal decomposition of the metric, $\gamma\ij = \frac 1\chi \tgm\ij$. The lapse and shift gauge parameters are given by $\alpha$ and $\beta^i$, {respectively}. 
In this section, $\chi$ is the metric conformal factor which relates to the cosmological scale factor as $\chi = 1/a^{2}$.  
The extrinsic curvature $K\ij$ is also split into its conformal traceless part $\tA\ij$ and the trace $K$, 
\bea
K\ij = \frac1\chi \left( \tA\ij +\frac13\tgm\ij K\right)~.
\eea
It relates with the Hubble rate $H$, in the homogeneous case, as
\bea
 H = - \frac { K} 3 ~ .
\eea

The energy-momentum tensor can be decomposed into the scalar fields' energy density $\rho_{\rm sf}$, momentum density $S_i$ and anisotropic tensor $S_{ij}$, 
\bea \label{3+1sources}
 \rho_{\rm sf}&=& n^\mu n^\nu T_{\mu\nu} ~,\\
 S_i &=& -\gamma^{\mu}_i n^\nu T_{\mu\nu} ~,\\ 
 S_{ij} &=& \gamma^{\mu}_i \gamma^{\nu}_j T_{\mu\nu} ~,\\ 
 S &=& \gamma\IJ S\ij ~,
\eea
where $n^\mu=(1/\alpha, -\beta^i/\alpha)$ is the unit normal vector to the three-dimensional slices. In analogy to the perfect fluid case with pressure $p =S/3$, the effective equation of state can be defined by 
\bea 
 \omega \equiv \frac{p}{\rho_{\rm sf}}= \frac 13 \frac{S}{\rho_{\rm sf}}~.
\eea

In the gravity sector, the energy associated with gravitational vector and tensor modes is given by 
\bea
  \rho_{\rm shear} = \frac {\Mpl^2}{2} \tilde A\ij \tilde A\IJ \propto \partial_t\tgm\ij \partial_t\tgm\IJ ~,
\eea
and the curvature contribution to the energy budged is written in terms of the Ricci scalar (of the 3-dim metric) 
\bea  
\rho_R = \frac{\Mpl^2}{2} R ~.
\eea

Then one can write the Hamiltonian and momentum constraint equations as 
\begin{align}
\mathcal{H} & =  \frac{\Mpl^2}{3} K^2 + \frac{\Mpl^2}{2} R  - \frac{\Mpl^2}{2} \tilde A_{ij}\tilde A^{ij} - \rho_{\rm sf}, \label{eqn:Ham} \\
  & = 3\Mpl^2 H^2 + \rho_R - \rho_{\rm shear} - \rho_{\rm sf}= 0\, , \nonumber \\[1mm] 
\mathcal{M}_i & = D^j (K_{ij} - \gamma_{ij} K) - 8\pi S_i =0\, . \label{eqn:Mom}
\end{align}

From the  Arnowitt-Deser-Misner formalism, it can also be shown that the conditions to have an accelerated expansion of the universe are given when 
\begin{gather}
\omega < -\frac13 
~, \qquad 
\rho_{\rm shear} < \left| \frac{3\rho_{\rm sf}} 4 \left( \frac13 + \omega \right) \right| ~.
\label{eq:inf_cond} 
\end{gather}

Averaging overall space, we can use these conditions to determine the beginning of inflation after the preinflationary era, as well as to set the time of which preheating starts.
\\

In the following analyses, the mean value of variable at a given time is denoted with $\langle ... \rangle$ brackets. {For instance, for a given variable $\theta$ }
\be \label{eqn:Kini}
\langle \theta \rangle \equiv \frac{1}{{\cal V}} \int \theta \, \rr d {\cal V}~,
\ee
where ${\cal V}$ is the spatial volume. Similarly, the root-mean-square (rms) and the standard deviations (std)  are computed like 
\be
{\rm rms} (\theta) = \sqrt{\langle \theta^2 \rangle}
~, \qquad 
{\rm std}(\theta) = \sqrt{\langle \theta^2 \rangle - \langle \theta \rangle^2 } ~,
\ee
These identities are used to assess the level of inhomogeneity in variable $\theta$, as well as the scope of local overdensities. Some example includes the density contrast $\delta_{\rho_{\rm sf}}$ and curvature contrast $\delta_{R}$  which are  given by 
\be \label{eq:deltarho}
\delta_{\rho_{\rm sf}}= \frac{ \rho_{\rm sf}- \langle \rho_{\rm sf}\rangle}{3\Mpl^2 H^2} 
~, \qquad 
\delta_{R}= \frac{ \rho_{R}- \langle \rho_{R}\rangle}{3\Mpl^2 H^2} ~.
\ee
The scalar curvature $\zeta$, as well as the mean number of efolds $\langle N \rangle $ are computed with 
\beq
 \langle N \rangle  = \langle \ln(a) \rangle  
 ~, \qquad 
 \zeta = {\rm std}\left[\ln(a)\right] ~,
\eeq
with $a = 1/\sqrt{\chi} $. Here, $\chi$ denotes the metric conformal factor in Eq.~(\ref{timeline}).
\\

\begin{figure*}[!t]
\begin{center}
\hspace{1cm} \textbf{Evolution in $\varphi$ and $\chi$} \hspace{4cm} \textbf{Evolution in $h$ and $s$}
\includegraphics[width=16.cm]{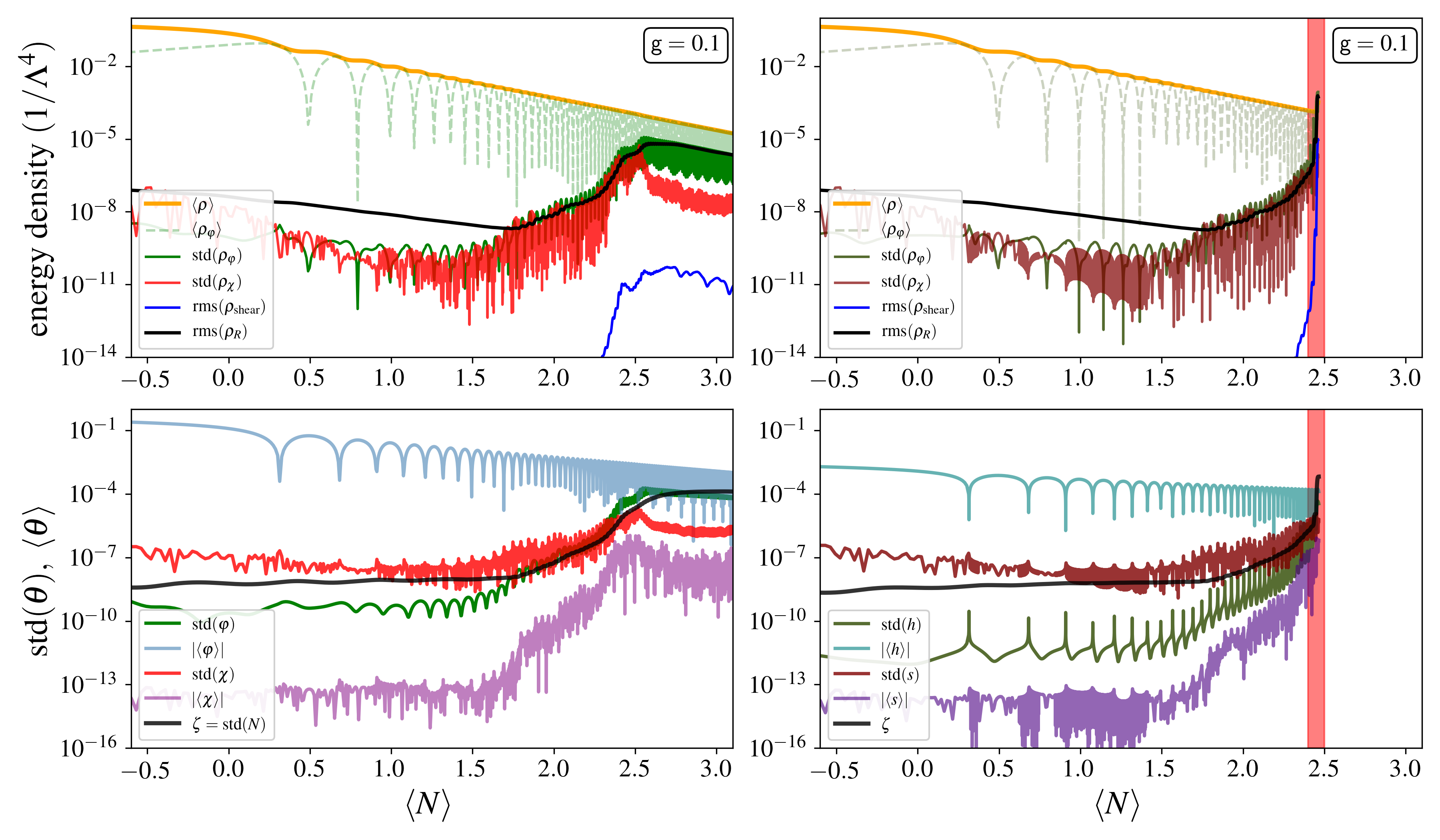}
\end{center}
\vspace*{-0.3cm}
 \caption{  
 Comparison of the two evolution schemes; Left panels use the canonical Einstein field $\varphi, \chi$ during the numerical evolution, right panels uses the Jordan-frame field $h, s$. The top panels show the time evolution of the standard deviations for the kinetic inflaton energy $\rho_\varphi$ (green line), auxiliary field kinetic energy $\rho_\chi$ (red line), and the root-mean-square values for the gravitational shear  $\rho_{\rm shear}$ (blue line), and curvature contributions $\rho_{\rm R}$ (black line). In both plots the mean energy densities, $\langle\rho_{\rm sf}\rangle $ (in orange) and $\langle\rho_{\varphi}\rangle $ (in dotted green line) have been added as a  reference.  The bottom panels show the evolution of scalar curvature $\zeta$ (in black) and the std  (in green and red lines) and mean values (blue and purple lines, respectively) of the scalar fields in the Jordan (left panel) and Einstein (right panel) frame notation. The fields are represented in units of Planck mass ($m_p$). Red shaded area indicates the region when dynamical instabilities in the $h$-field evolution raises large  violations in the constraint equations (\ref{eqn:Ham}).  
 \label{fig:reheating}
}
\end{figure*}

\subsection{Computational details \label{sec:numdetails} }

All simulations are done  in a grid composed by $(128)^3$ to $(156)^3$ cells with an initial grid-size $L$ which is of the order of the Hubble size. The topology is of a 3-dimensional torus with periodic boundary conditions in all dimensions. The initial configurations assume conformal flatness, (e.g. $\tilde\gamma\ij = {\rm diag}(1,1,1)$ and $\tilde A\ij = 0$),
where inhomogeneities are contained in the form of scalar field gradients, which are then compensated by the conformal factor (i.e. gravitational scalar curvature). The valid sets of initial data have been computed by solving the Hamiltonian constraint iteratively, as in most of the previous works \cite{Clough_2015,Aurrekoetxea_2020,Joana2020,Garfinkle_2020}.  The evolution of the system is computed in the Einstein frame by numerical integration of the BSSN equations \cite{PhysRevD.52.5428,Baumgarte_1998,10.1143/PTPS.90.1} in 3+1 dimensions, implemented in the GRChombo code. A more detailed explanation of the structure and validation of the code can be found in the appendixes and in Refs.\cite{Clough_2015,Andrade2021}. 
\\

\begin{figure*}[!t]
\begin{center}
\includegraphics[width=16.cm]{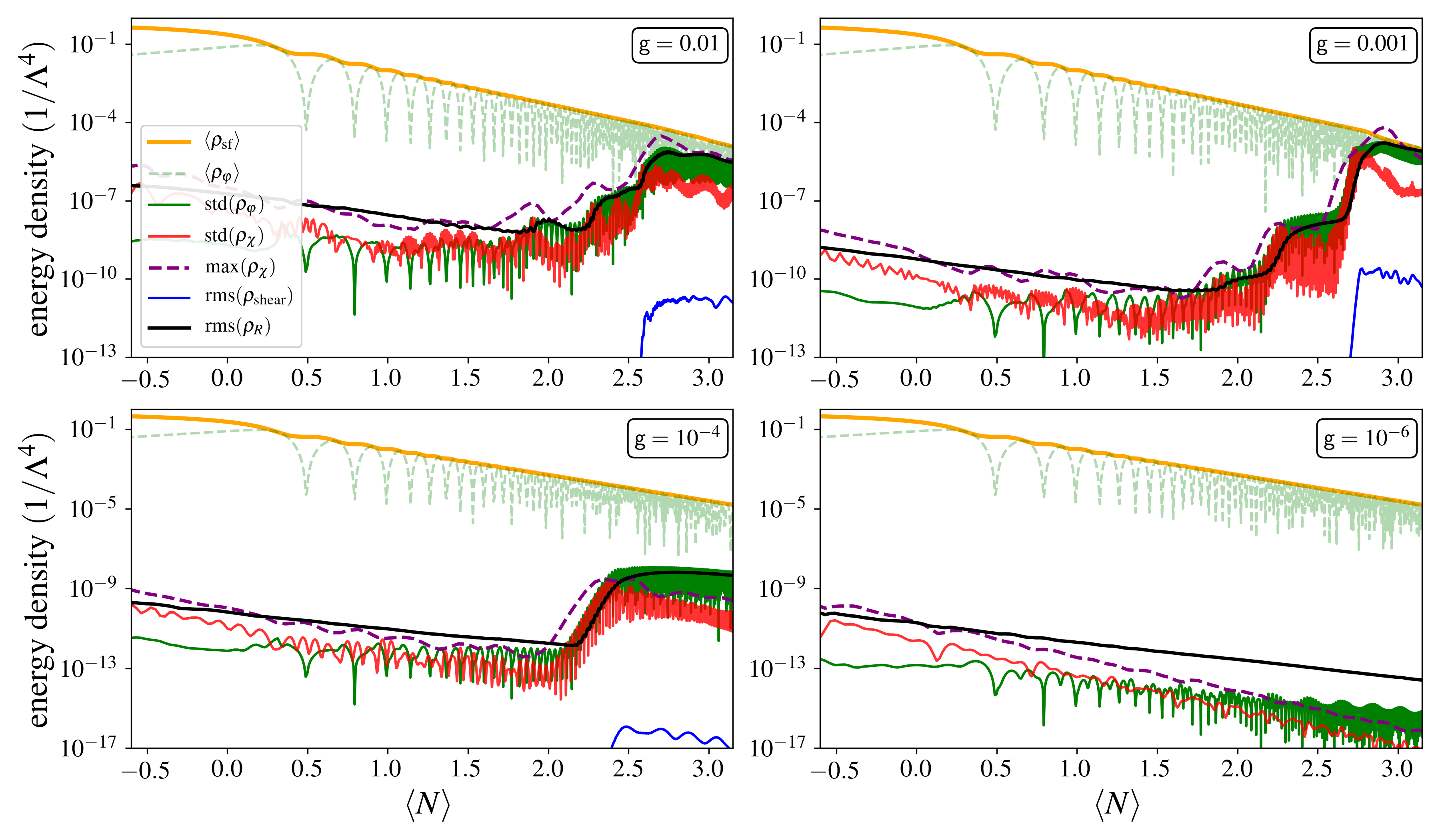}
\end{center}
\vspace*{-0.5cm}
\caption{  \label{fig:reheating2}
Same as in the top left panel in figure \ref{fig:reheating} for several simulations with different choices in $\mathsf{g}$. Dashed purple lines denote the maximum values of the auxiliary field's kinetic energy. The box size of the simulations at the end of inflation correspond to $L\approx 2 H^{-1}$.
}
\end{figure*}

Two different evolution schemes have been used for the numerical evolution of the fields. Simulations on the preheating epoch are evolved using the canonical Einstein fields $\varphi,\> \chi$. Therefore, at each timestep,  the approximate conversions of Eq.~(\ref{eq:approx_fieldconvers}) are used to recover the values of $h,\> s$ needed to evaluate the potential, Eq.~(\ref{eq:V_HIaux}), and its derivatives.  This is done to solve dynamical instabilities occurring at $h\approx 0$, where the $h$-field experience transients accelerations as a result of the presence of $\Gamma^h_{hh}$-term in the evolution equations Eq.~(\ref{eq:eomsf_J}) (see also Fig.~\ref{fig:Higgs_EJ}). Despite that this issue can be overcome by shortening the time integration during the coherent linear phase, the code becomes very unstable during the broad resonance period, when the $h$-field inhomogenizes.  This issue is solved when the system is evolved using the Einstein-frame notation, and it allows us to continue the simulations for a longer time. As shown in Sec.~\ref{sec:SimsReheat}, both evolution schemes give numerically equivalent results. 
On the other hand, simulations on the preinflationary era are done using the (exact) formalism with the Jordan-framed  $h$ and $s$ fields. This does not represent an issue, as $h$ does not continuously oscillate around zero and therefore the instability is not present.

\section{Dynamics of preheating \label{sec:SimsReheat} } 

At the end of inflation, the inflaton field starts a period of coherent oscillations around the potential minimum. The large amplitude of the oscillations justifies a classical treatment of the inflaton field. The simulations start about $N_{\rm ini} \approx -1$ $e$-folds before the end of inflation, thus the field is considered to be initially homogeneous\footnote{
Simulations containing initial perturbations in the Higgs field has also been considered without significant changes in the resonance dynamics. See the Appendix \ref{App:AuxiliarlyFigs}. 
} 
in field value at the edge of the plateau, and is rolling down the potential with a background kinetic term. The initial values are set to
\beq
h \approx  1.1 \cdot 10^{-2} 
~, \qquad 
\Pi_h  \approx - 8.1 \cdot 10^{-9} ~\Mpl 
\eeq
which is equivalent to 
\beq
\varphi \approx 2.0 ~\Mpl 
~, \qquad 
\Pi_\varphi \approx -1.2 \cdot 10^{-6} ~\Mpl^2  
\eeq
where $\Pi_h,$ and $\Pi_\varphi$ correspond to the fields' momentum. \\

On the other hand, the auxiliary field is assumed to be in its vacuum state due to the redshift caused during inflation, where fluctuations of the field are of quantum origin. The initial state of the field is set by
\beq
\begin{split}
s(\vec x) = \ &\langle s_0 \rangle +
\sum_{n=1}^{N_m}\sum_{i=1}^3\frac{\Delta_{n}}3  \cos  \left( \frac{2\pi n x}\lambda   + \theta_n \right) 
\\
&\text{with  } ~ \langle s_0 \rangle = 0~, \quad \Delta_{n} = \frac{\pi n}\lambda
\end{split}
\eeq
where $\lambda \approx L/10$ is the largest perturbation size,  $\theta_n$ is a random phase, and the number of modes $N_m$ is set between $10$ and $50$.  
The momentum of the $s$-field (and $\Pi_\chi$) is initially set to zero.

\subsection{Parametric resonances}

In the analysis, the evolution of the gravitational  and scalar field sector are considered, i.e. $\rho_{\rm shear},\> \rho_R,\>\text{and } \rho_{\rm sf}$. The later one is further decomposed into the inflaton and auxiliary field parts, $\rho_\varphi,\> \rho_\chi$ by assuming
\beq      
   \rho_\varphi \approx \frac12 \Pi^2_\varphi       
   ~, \qquad 
   \rho_\chi \approx \frac12 \Pi^2_\chi       
\eeq
where the kinetic energy of the fields is used as a proxy to estimate their total energy contribution.

Because we can use two evolution schemes for the fields, namely using the Jordan or the Einstein frame definitions,  let us first compare both schemes and ensure that we obtain equivalent results. In Fig. \ref{fig:reheating}, this is done by direct comparison of the evolution of the energy densities shown in the top panels.
The mean energy density is shown with the orange lines,  while the standard deviation of the fields kinetic energies $\rho_\varphi$ and $\rho_\chi$, corresponding to the green and red lines, respectively. Additionally, the densities from the gravitational curvature (black lines) and shear (blue lines) are also shown.
In the bottom panels, the mean and standard deviation of the fields and scalar curvature $\zeta$ are shown. It is interesting to note several differences in the evolution of the fields: Because the shape of the potential is different in both representations of the field, this is reflected in the scaling of mean values of the fields \cite{Martin2010}. In particular, these simulations show that the mean of Higgs field scales like  $\langle h\rangle \propto a^{-3/4}$. On the other hand, the canonically normalized inflaton scales like $\langle\varphi\rangle \propto a^{-3/2}$. The latter is analogous to a quadratic potential around its minimum \cite{Martin2010,Giblin2019}. Differences are also noticeable when looking at the field excitations (standard deviations): in the evolution of perturbations in $h$, the Riemann spikes that occur when $h\approx 0$ are clearly visible, while for perturbations in $\varphi$ they are hidden because of the mixing with $R$, in the Einstein frame. In both cases, though, the scalar perturbation $\zeta$ closely follow the fluctuations of the  $h,\varphi$ fields. On the other hand, the auxiliary fields behave very similarly in both schemes, and we can clearly relate the $s$ and $\chi$ fields. This is not surprising as the $s$-field is chosen to be minimally coupled to gravity and therefore the mixing after the Weyl transformation is predominantly between $h$ and $R$. All in all, we see that, for this particular case, the broad resonance of the fields occurs within $N \approx 1.5 - 2.5$ efolds, where the excitations of the fields (in all frames) grow exponentially.

\begin{figure}[b!]
\begin{center}
\hspace*{-0.5cm}
\includegraphics[width=8.cm]{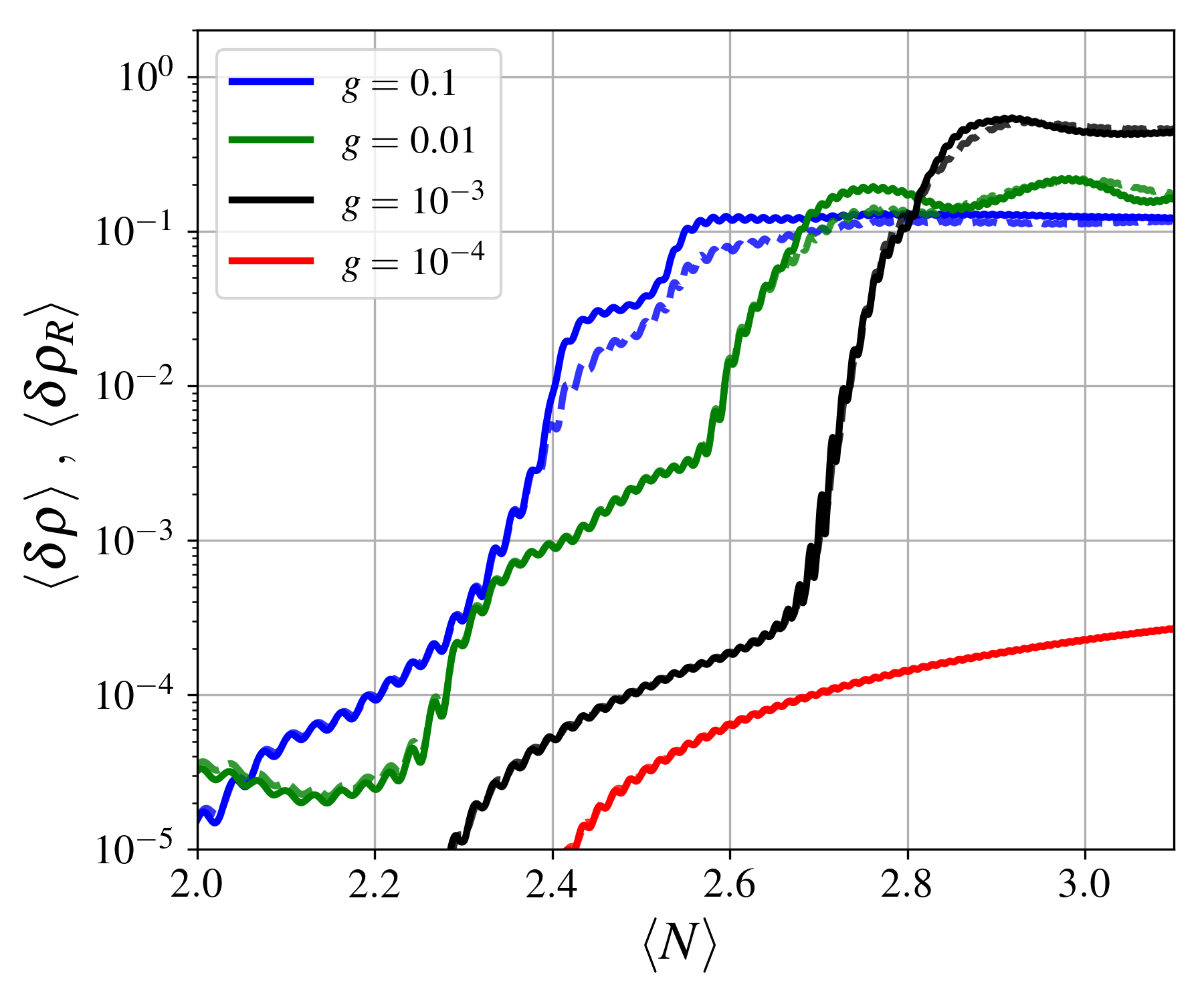} 
\includegraphics[width=8.5cm]{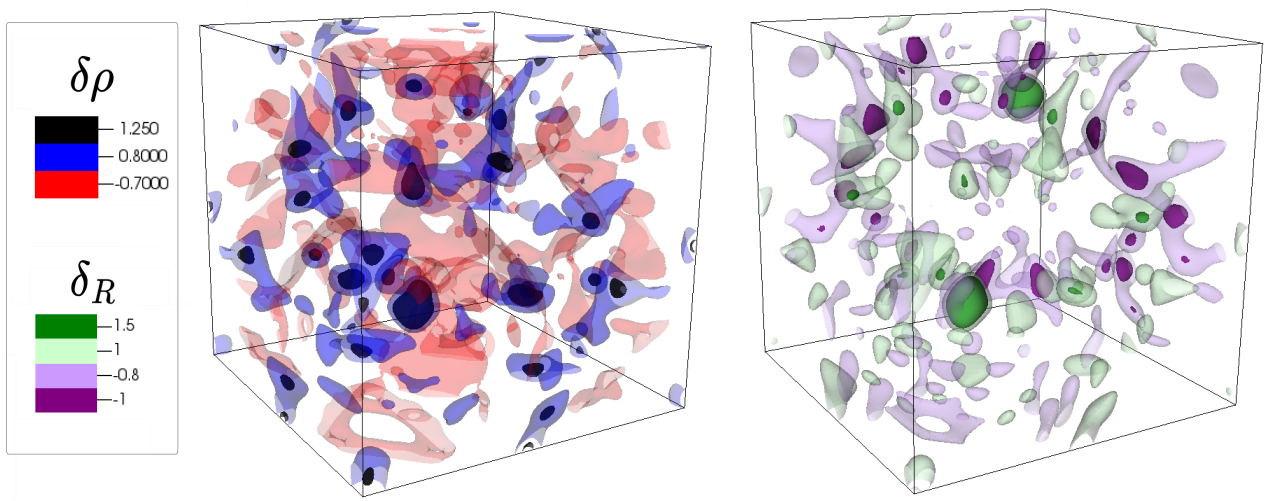}
\end{center}
\vspace*{-0.5cm}
\caption{  \label{fig:reh_structure}
Structure formation during reheating. Top panel shows the evolution of the global density contrasts for $\rho_{\rm sf}$ (solid line) and $\rho_R$ (dashed lines). Bottom plots show contours of under/overdense regions: $\rho_{\rm sf} = 1.25$ (black), $\rho_{\rm sf} = 0.8$ (blue), and $\rho_{\rm sf} = -0.7$ (red), as well for negative/positive curvature: $\rho_{R} = 1.5$ (dark green), $\rho_{R} = 1.0$ (light green), $-0.8$ (light purple), and $\rho_{R} = -1.0$ (dark purple).  These 3D representations correspond to the simulation with $\mathsf{g} = 0.001$ shown above,  at $N \approx 2.9$. 
}
\end{figure}

In a similar way, {Fig.$\>$\ref{fig:reheating2}} shows simulations for different values in $\mathsf{g}$. During the broad resonance period, we find that curvature grows strictly following the excitations of the fields (the std values). The efficiency of resonances is conditioned by the $\mathsf{g}$-coupling, as the interaction term (Eq.~\ref{eq:L_int}) can be interpreted as the effective mass terms of the fields. 
For large couplings,  $\mathsf{g} \gtrsim 1$, fluctuations of the field are largely suppressed during the last efolds of inflation, pushing the field down to zero. This overdamping, which is partly due to  the classical treatment of the initial gradients, impedes the resonance periods at  later times and preheating fails.  Coupling strengths in the range of  $ 0.1 \lesssim\mathsf{g} \lesssim 10^{-3}$ does allow preheating. However, while at larger coupling values the broad resonant phase earlier, the produced fluctuations saturate at lower energies resulting in lower energy transfer from the background field.   

For even lower coupling values, $\mathsf{g} \lesssim 10^{-4}$, the particle production  becomes inefficient  (at least during the first 3-4 efolds post-inflation) and the energies associated with them fail to co-dominate the dynamics. 
The preheating of the universe is therefore presumably  delayed to later times, but we cannot numerically explore this region.  

In summary, within the assumptions of the model, a successful and fast preheating of the Universe occurs for a range in the field-field strength coupling of $ 1 \gtrsim \mathsf{g} \gtrsim 10^{-4}$, with a peak efficiency of around $ \mathsf{g}  \approx 10^{-3}$. This is a surprising result because other studies on Higgs inflation \cite{Sfakianakis2019} found that self-resonances from the Higgs, alone, effectively preheat the universe when considering linearized gravity. These simulations show that this is no longer the case when considering full gravity. 

\subsection{Structure formation}

Structure formation starts when the energy fluctuations of the fields grow comparable to the background energy density. In our simulations, this occurs around $N \gtrsim 2.5$ efolds after the end of inflation, flagging the highly non-linear phase in both matter and gravitational sectors.  As shown in Fig. \ref{fig:reh_structure}, the structure consists of the region of space containing both under- and  overdense energies. Overdense (underdense) scalar-field regions coexist with large local positive (negative) Ricci scalar fluctuations of the order of $\delta\rho, \delta\rho_R \approx 1$ ($\delta\rho, \delta\rho_R \approx -1$), reaching even larger values for low-mass particles (i.e. $  \mathsf{g} \approx 10^{-3} $). The type of structure formed in these simulations resembles to what was reported in other works as oscillons (or \textit{transfers} \cite{Lozanov2019}). 
Because during the structure formation the dominant energies are shifted to smaller scales, our simulations can not accurately run long enough to confirm the formation of black holes. However, other works have shown that instabilities on such oscillon-like objects, can lead to the formation of primordial black holes through self-collapse \cite{Kou:2019bbc,Nazari2021,de_Jong_2022}.  This will be studied with dedicated simulations in future works.

\section{Dynamics of preinflation \label{sec:SimsPreinf}}

\begin{figure*}[t!]
\begin{center}
\hspace{0.5cm} \textbf{Super-Hubble perturbations } \hspace{3.4cm} \textbf{Sub-Hubble perturbations }
\includegraphics[width=8.5cm]{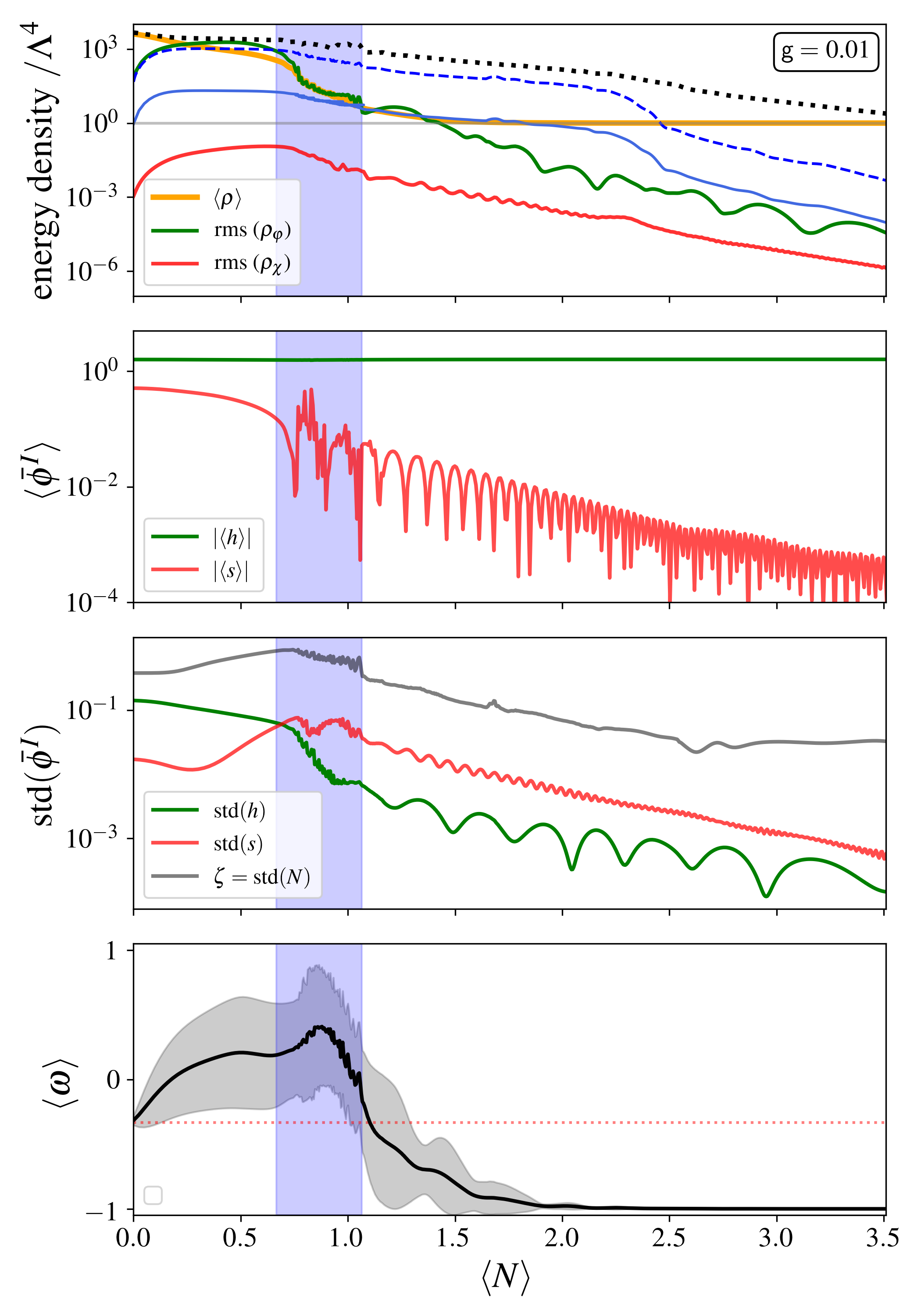}
\includegraphics[width=8.5cm]{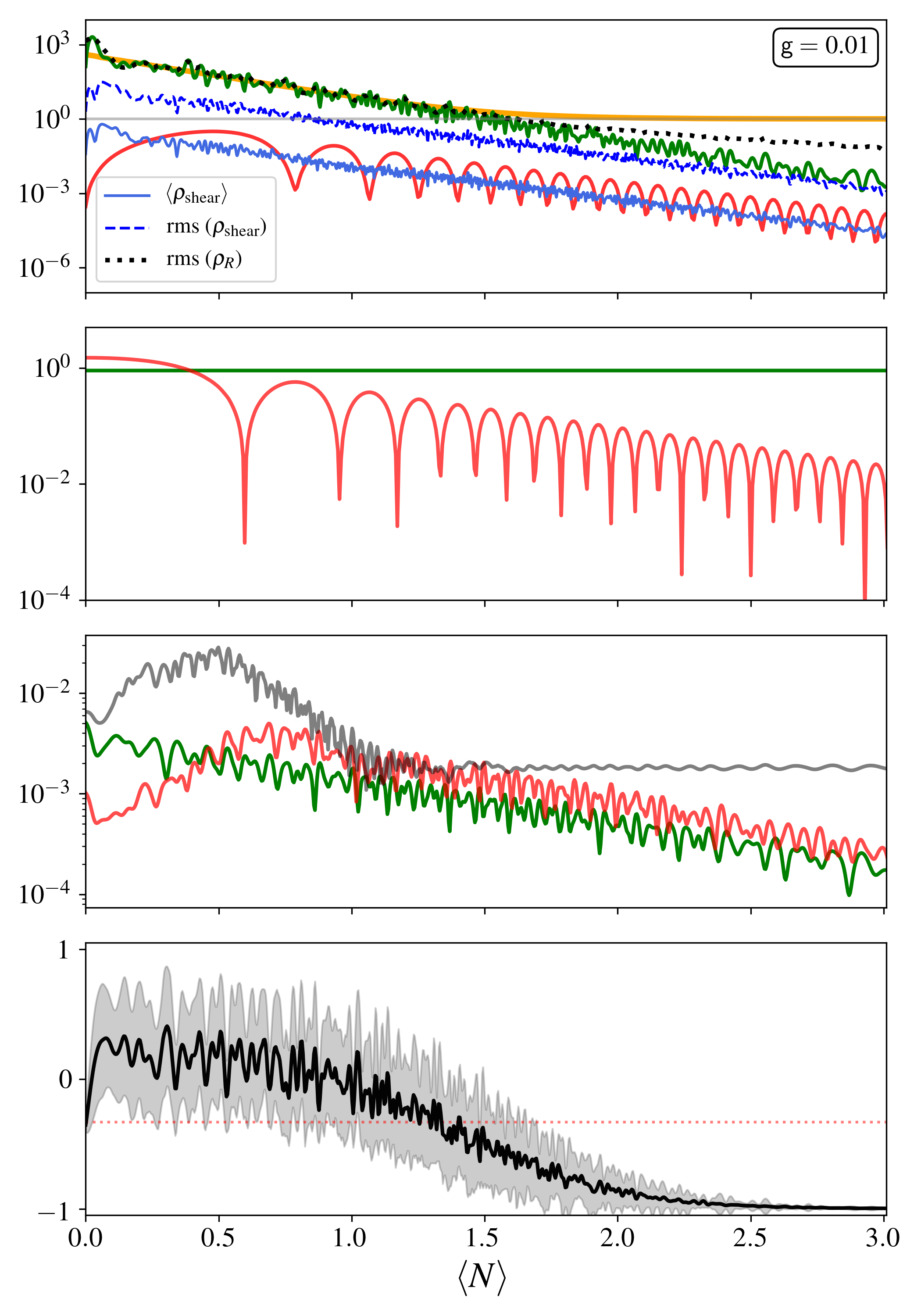}
\caption{ \label{fig:preinflation} 
Dynamics of two example simulations initially at super-Hubble (left panels) and sub-Hubble (right panels) perturbations. Top panels show the evolution of scalar field's energy density (orange line), and shear (solid blue line). The rms values for the Higgs (green line), auxiliary field (red line) as well as for the gravitational shear (dotted blue line) and curvature densities (dotted black lines). The mean field evolution (upper-middle panels) and std values (lower-middle panels) for the Higgs (green line) and auxiliary (red line) fields. The scalar curvature perturbation $\zeta$ is shown in gray lines. Bottom panels show the evolution of the equation of state (solid black line), with plus/minus std values in the shaded gray area. The red-dotted line denotes the $\omega < -1/3$ threshold necessary for accelerated expansion of Universe.   The initial box size of the simulations correspond to $L\approx 5 H^{-1}$.
}
\end{center}
\end{figure*}

The preinflationary scenario is dependent on the initial conditions of the universe, and therefore, its properties are unknown. Arguably, in the classical regime, the primordial universe can be thought of as an inhomogeneous inflaton field that successfully leads to inflation when the kinetic and gradient energies fall below the field's potential energy. The necessary conditions to trigger exponential expansion are a negative effective equation of state that $\langle \omega \rangle < -1/3$, and a subdominant contribution of gravitational modes, i.e. Eq.~(\ref{eq:inf_cond}). If these conditions are satisfied quickly enough, so that the mean field values are still in the flat part of the potential, then inflation starts. 
\\

The particular case of (single field) Higgs inflation model was considered in our previous paper, Ref.~\cite{Joana2020}. The model showed to be robust to large inhomogeneities at sub- and super-Hubble scales. Our simulations showed that highly dynamical field fluctuations source large gravitational (shear and tensor) modes that can eventually dominate the energy budget. The energy density associated with field fluctuations decays like radiation, $\rho_{\rm sf}\propto a^{-4}$, and these gravitational modes like 
$\rho_{\rm shear} \propto a^{-2}$. In any case, both scalar-field and gravitational excitations eventually become subdominant in just a few $e$-folds and inflation begins. In the following, these analyses are expanded by adding an auxiliary field. 
\\

The simulations on preinflation initially contain field gradients in  both the inflaton and auxiliary fields, with perturbation in sub- and super-Hubble configurations. The initial mean value of the Higgs (inflaton) is always considered to be beyond $ \langle h \rangle  > 0.5$,  $(\langle \varphi \rangle /\Mpl > 6)$, so it is deeply located in the flat region of the potential. For the auxiliary field, cases with zero and non-zero mean values have been considered. The selection of these cases have been chosen so that the overall mean energy density is a few orders of magnitude larger than the energy scale of inflation, i.e. $\langle \rho_{\rm sf}(t_0) \rangle >  \Lambda$. Thus, all considered cases contain inhomogeneities well beyond the linear regime. 
\\

Figure \ref{fig:preinflation} shows two example cases, at super-Hubble (left) and sub-Hubble (right) scales. In both cases, the preinflationary phase consists of a homogenization period driven by the (in average) positive expansion of the Universe. Similarly as shown in Ref.~\cite{Joana2020}, super-Hubble initial conditions tend to form trapped surfaces, or pre-inflationary black holes (PIBHs), after Hubble crossing. Because these black holes are  always (much) smaller than the Hubble radius,  instead of impeding inflation, they tend to facilitate it by trapping the overdense regions, thereby fastening the homogenization. On the other hand, at sub-Hubble scales, perturbation modes transit back-and-forth between gradients and kinetic energies, effectively making the energy density scale like radiation $\rho_{\rm sf}\propto a^{-4}$. In that scenario, the optimal conditions for triggering inflation look like a dynamical attractor, and cosmic inflation starts within a few efolds.  
\\

Interestingly, in the presence of the auxiliary (spectator) field,  these oscillations also trigger energy transfer between the (Jordan framed-) Higgs and the auxiliary field. 
However, these dynamics do not originate in enhancement of structures like in preheating, because now at field values $ h > 0.02$, the non-minimal coupling of the Higgs has the effect of significantly reducing the impact of the (minimally coupled) auxiliary field when seen in the Einstein frame. This effect can be observed in Fig. \ref{fig:preinflation}, where even when perturbations in the $s$-field are larger than in the $h$-field (see Fig.~\ref{fig:preinflation} middle panels), the dynamical term is always orders of magnitude smaller in the auxiliary field $\chi$.  
The suppression effect comes from the mixing, which is introduced by the field-space metric lower than unity, i.e. ${\cal G}_{ss} = 1/({1+\zeta_h h^2}) \leq 1$ and becomes orders of magnitude smaller for large enough $h$-field values (e.g. ${\cal G}_{ss}\ll 1$ for $h > 0.02$). 
This suppression factor should apply to all other possible matter components that are minimally coupled to gravity.

\subsection{On the initial conditions for inflation \label{sec:OnInitialConditions}}

In this paper we have extended previous works on testing the initial conditions for inflation by including the interplay of an extra (minimally coupled) scalar field. We have tested cases when the initial configuration of the inflation is deeply inhomogeneous but with its mean-field value inside the slow-roll region of the potential.  In the context of Higgs inflation, this corresponds to a mean value close to the plateau. Initial states where the mean-field is in the non-inflationary region (i.e. in the bottom of the potential) have not been considered as these cases should not lead to inflationary regions, as tested in Ref.~\cite{Joana2020}. This is because gradients terms make the fields oscillate around the mean value, thus these inhomogeneities are not capable  of driving the field up to the plateau. One could still consider large field inhomogeneities which spans the scalar field into the potential's plateau, however these perturbations are necessarily super-Hubble (at sub-Planckian gradient energies) and, thus, these regions can be treated as separate universes. This is  particular to Higgs inflation, as the plateau starts at $\varphi > \Mpl$. 
\\

Our initial settings have also assumed a conformally flat expanding universe. These scenarios corresponds to the case with only scalar perturbations in the gravitational part, (i.e. without vector and tensor gravitational modes). This is  related to the choice of considering a null kinetic term in the initial hypersurface, which can be seen as a rather ``special" slicing choice at the instantaneous initial time  where scalar-field kinetic terms have been gauged away, trivially satisfying the momentum constraint Eq.~(\ref{eqn:Mom}). Because these initial setting are highly dynamical, this kind of slicing is not stable and once the system is time-evolved both gravitational modes and scalar field kinetic terms are quickly generated, leading to a less symmetric inhomogeneous system. In particular, one could have chosen an analogous situation with initially homogeneous field values but with largely inhomogeneous kinetic terms which would raise scalar-field inhomogeneities in the immediate time evolution \cite{Joana2020}. Because the minimally coupled fields are energetically subdominant at high enough Higgs values, the previous picture still holds beyond the single-field case. 
\\

Nonetheless, there are still several limitations with such initial settings. The assumption on conformal flatness only allow for gravitational perturbations risen by the scalar field evolution, and therefore independent large tensor metric perturbation are ignored. Studying these cases requires solving (non-trivially) both the Hamiltonian and momentum constraints and future works will deal with this challenge. In addition, this work has assumed that only the Higgs field has a non-minimal coupling to gravity, serving as a reference for other more specific models like quintessential Higgs inflation \cite{Es-haghi:2020oab}, two Higgs doublet models \cite{Lee:2021rzy}, etc. 
Still, systems with two or more non-minimally coupled fields can show a much richer dynamical evolution during (pre-)inflation; Exploring complex trajectories in field space and possibly including multiple inflationary phases at distinct energy stages. All these considerations are left to future works.
\\

Under the previous considerations, in all the considered cases, we find common dynamical patterns of the Higgs pre-inflationary era. This phase can be described as a homogenization era with a varying inhomogeneous equation of state, which effectively  correspond to a radiation dominated universe $\langle \omega \rangle \approx 1/3$. 
Once the scalar-field falls below the energy scale of the inflationary potential, the equation of state tends to a de-Sitter Universe with $\omega \approx -1$, satisfying the first condition for inflation, i.e. Eq.~(\ref{eq:inf_cond}). It has also been shown that strong field dynamics near the Hubble scales develop large gravitational modes that potentially influence the expansion of the Universe until they become subdominant. These modes effectively delay the beginning of inflation, but do not prevent it. Moreover, during the pre-inflationary era, lasting $N \approx 3\text{-}7$ efolds, the variation on the average value of the inflaton $\varphi$ is negligible, which prevents the ``overshooting'' problem seen in other models \cite{Aurrekoetxea_2020}. 
All these considerations make me conclude that, under the considered settings, the Higgs inflation model is very robust to the inhomogeneous multi-field initial conditions of the pre-inflationary era.

\section{Conclusions \label{sec:Conclusion}}

In this paper, I have used fully general relativistic simulations to investigate the robustness of the Higgs inflation model to inhomogeneous multi-field initial conditions. Specifically, in the presence of additional field couplings, these being necessary for a parametric-type reheating. 
It is shown that, at large enough Higgs values,  the non-minimal coupling of the Higgs protects the dynamics of the inflaton by diminishing the impact of couplings to other fields and matter sectors.  And, as shown in Ref.~\cite{Joana2020}, the dynamics from gravitational shear and tensor modes can only delay, but not prevent, cosmic inflation.
\\

Additionally, simulations on the preheating dynamics of the two-field system were presented where full gravitational backreactions in the metric have been considered. As expected, it is shown that the efficiency of the preheating is conditioned to coupling strength between the fields. In particular, for such a simple model, it was found the preheating of the Universe within the first $3-4$ efolds post-inflation to occur for couplings in the range of $ 0.1\lesssim \mathsf{g} \lesssim 10^{-4}$. On the other hand, self-resonance from the Higgs alone, fails to reheat the Universe within the first $3-4$ efolds after inflation. 
\\

These simulations have also shown the formation of complex structures during the preheating, consisting in large under/overdensities as well as strong positive/negative (local) curvature regions, suggesting the possibility of (seeding) later formation of compact structures like primordial black holes. Nonetheless, these results should be taken cautiously as further investigations, including dedicated numerical simulations, are necessary to accurately resolve these highly non-linear objects. 
\\

Future works are also necessary to study more realistic preheating scenarios, including the Higgs couplings to the Standard Model particles. These are important, because they potentially could shorten the preheating within one $e$fold after inflation \cite{Ema2017,Ema2021}, if metric backreactions allow it. Other interesting aspects to be studied are the emission of gravitational waves  - and possible amplification effects\cite{Chunshan2016,Lozanov2019,Zihan2020,Cai2021}. Importantly, some of the described  phenomena are expected to be in the observable range of future gravitational-wave experiments. 
\\

\section{Acknowledgments}

The author warmly thanks Christophe Ringeval and Sebastien Clesse for helpful comments and support, as well as Pierre Auclair, Tiago França, Miren Radia, Josu Aurrekoetxea, Katy Clough and Eugene Lim for useful discussions. He is also in gratitude to all members of the \texttt{GRChombo} team (https://www.grchombo.org/\#people) for their work and maintenance on the code. 
This work is supported by the {FRIA} Grant No.1.E.070.19F of the Belgian Fund for Research, F.R.S.-FNRS. 
Computational resources have been provided by 
the Julich Supercomputing Center JUWELS HPC under PRACE grant Tier-0 Proposal No. 202022535, 
the Consortium des Équipements de Calcul Intensif (CÉCI) under Grant No. 2.5020.11 by the Walloon Region. Parallel code developments were done on the CURL cosmo clusters at UCLouvain, funded by the “Fonds de la Recherche Scientiﬁque - FNRS” under Grant No. T.0198.19. 
Analysis and visualizations employed the Visit \cite{10.5555/2422936} and yt-project \cite{Turk_2010} software packages.

\bibliographystyle{apsrev}
\bibliography{biblio.bib}

\begin{thebibliography}{86}
\expandafter\ifx\csname natexlab\endcsname\relax\def\natexlab#1{#1}\fi
\expandafter\ifx\csname bibnamefont\endcsname\relax
  \def\bibnamefont#1{#1}\fi
\expandafter\ifx\csname bibfnamefont\endcsname\relax
  \def\bibfnamefont#1{#1}\fi
\expandafter\ifx\csname citenamefont\endcsname\relax
  \def\citenamefont#1{#1}\fi
\expandafter\ifx\csname url\endcsname\relax
  \def\url#1{\texttt{#1}}\fi
\expandafter\ifx\csname urlprefix\endcsname\relax\def\urlprefix{URL }\fi
\providecommand{\bibinfo}[2]{#2}
\providecommand{\eprint}[2][]{\url{#2}}

\bibitem[{\citenamefont{Starobinsky}(1980)}]{STAROBINSKY198099}
\bibinfo{author}{\bibfnamefont{A.}~\bibnamefont{Starobinsky}},
  \bibinfo{journal}{Physics Letters B} \textbf{\bibinfo{volume}{91}},
  \bibinfo{pages}{99} (\bibinfo{year}{1980}), ISSN \bibinfo{issn}{0370-2693},
  \urlprefix\url{https://www.sciencedirect.com/science/article/pii/037026938090670X}.

\bibitem[{\citenamefont{Guth}(1981)}]{PhysRevD.23.347}
\bibinfo{author}{\bibfnamefont{A.~H.} \bibnamefont{Guth}},
  \bibinfo{journal}{Phys. Rev. D} \textbf{\bibinfo{volume}{23}},
  \bibinfo{pages}{347} (\bibinfo{year}{1981}),
  \urlprefix\url{https://link.aps.org/doi/10.1103/PhysRevD.23.347}.

\bibitem[{\citenamefont{Sato}(1981)}]{10.1093/mnras/195.3.467}
\bibinfo{author}{\bibfnamefont{K.}~\bibnamefont{Sato}},
  \bibinfo{journal}{Monthly Notices of the Royal Astronomical Society}
  \textbf{\bibinfo{volume}{195}}, \bibinfo{pages}{467} (\bibinfo{year}{1981}),
  ISSN \bibinfo{issn}{0035-8711},
  \eprint{https://academic.oup.com/mnras/article-pdf/195/3/467/4065201/mnras195-0467.pdf},
  \urlprefix\url{https://doi.org/10.1093/mnras/195.3.467}.

\bibitem[{\citenamefont{Linde}(1982)}]{LINDE1982389}
\bibinfo{author}{\bibfnamefont{A.}~\bibnamefont{Linde}},
  \bibinfo{journal}{Physics Letters B} \textbf{\bibinfo{volume}{108}},
  \bibinfo{pages}{389} (\bibinfo{year}{1982}), ISSN \bibinfo{issn}{0370-2693},
  \urlprefix\url{https://www.sciencedirect.com/science/article/pii/0370269382912199}.

\bibitem[{\citenamefont{Akrami et~al.}(2020)\citenamefont{Akrami, Arroja,
  Ashdown, Aumont, Baccigalupi, Ballardini, Banday, Barreiro, Bartolo, and
  et~al.}}]{Akrami:2018odb}
\bibinfo{author}{\bibfnamefont{Y.}~\bibnamefont{Akrami}},
  \bibinfo{author}{\bibfnamefont{F.}~\bibnamefont{Arroja}},
  \bibinfo{author}{\bibfnamefont{M.}~\bibnamefont{Ashdown}},
  \bibinfo{author}{\bibfnamefont{J.}~\bibnamefont{Aumont}},
  \bibinfo{author}{\bibfnamefont{C.}~\bibnamefont{Baccigalupi}},
  \bibinfo{author}{\bibfnamefont{M.}~\bibnamefont{Ballardini}},
  \bibinfo{author}{\bibfnamefont{A.~J.} \bibnamefont{Banday}},
  \bibinfo{author}{\bibfnamefont{R.~B.} \bibnamefont{Barreiro}},
  \bibinfo{author}{\bibfnamefont{N.}~\bibnamefont{Bartolo}}, \bibnamefont{and}
  \bibinfo{author}{\bibnamefont{et~al.}}, \bibinfo{journal}{Astronomy \&
  Astrophysics} \textbf{\bibinfo{volume}{641}}, \bibinfo{pages}{A10}
  (\bibinfo{year}{2020}), ISSN \bibinfo{issn}{1432-0746},
  \urlprefix\url{http://dx.doi.org/10.1051/0004-6361/201833887}.

\bibitem[{\citenamefont{Ade et~al.}(2016)\citenamefont{Ade, Aghanim, Arnaud,
  Arroja, Ashdown, Aumont, Baccigalupi, Ballardini, Banday, and
  et~al.}}]{Ade:2015lrj}
\bibinfo{author}{\bibfnamefont{P.~A.~R.} \bibnamefont{Ade}},
  \bibinfo{author}{\bibfnamefont{N.}~\bibnamefont{Aghanim}},
  \bibinfo{author}{\bibfnamefont{M.}~\bibnamefont{Arnaud}},
  \bibinfo{author}{\bibfnamefont{F.}~\bibnamefont{Arroja}},
  \bibinfo{author}{\bibfnamefont{M.}~\bibnamefont{Ashdown}},
  \bibinfo{author}{\bibfnamefont{J.}~\bibnamefont{Aumont}},
  \bibinfo{author}{\bibfnamefont{C.}~\bibnamefont{Baccigalupi}},
  \bibinfo{author}{\bibfnamefont{M.}~\bibnamefont{Ballardini}},
  \bibinfo{author}{\bibfnamefont{A.~J.} \bibnamefont{Banday}},
  \bibnamefont{and} \bibinfo{author}{\bibnamefont{et~al.}},
  \bibinfo{journal}{Astronomy \& Astrophysics} \textbf{\bibinfo{volume}{594}},
  \bibinfo{pages}{A20} (\bibinfo{year}{2016}), ISSN \bibinfo{issn}{1432-0746},
  \urlprefix\url{http://dx.doi.org/10.1051/0004-6361/201525898}.

\bibitem[{\citenamefont{Goldwirth and Piran}(1990)}]{Goldwirth:1989pr}
\bibinfo{author}{\bibfnamefont{D.~S.} \bibnamefont{Goldwirth}}
  \bibnamefont{and} \bibinfo{author}{\bibfnamefont{T.}~\bibnamefont{Piran}},
  \bibinfo{journal}{Phys. Rev. Lett.} \textbf{\bibinfo{volume}{64}},
  \bibinfo{pages}{2852} (\bibinfo{year}{1990}).

\bibitem[{\citenamefont{Goldwirth}(1991)}]{Goldwirth:1990pm}
\bibinfo{author}{\bibfnamefont{D.~S.} \bibnamefont{Goldwirth}},
  \bibinfo{journal}{Phys. Rev.} \textbf{\bibinfo{volume}{D43}},
  \bibinfo{pages}{3204} (\bibinfo{year}{1991}).

\bibitem[{\citenamefont{Laguna et~al.}(1991)\citenamefont{Laguna, Kurki-Suonio,
  and Matzner}}]{Laguna:1991zs}
\bibinfo{author}{\bibfnamefont{P.}~\bibnamefont{Laguna}},
  \bibinfo{author}{\bibfnamefont{H.}~\bibnamefont{Kurki-Suonio}},
  \bibnamefont{and} \bibinfo{author}{\bibfnamefont{R.~A.}
  \bibnamefont{Matzner}}, \bibinfo{journal}{Phys. Rev.}
  \textbf{\bibinfo{volume}{D44}}, \bibinfo{pages}{3077} (\bibinfo{year}{1991}).

\bibitem[{\citenamefont{Kurki-Suonio et~al.}(1993)\citenamefont{Kurki-Suonio,
  Laguna, and Matzner}}]{KurkiSuonio:1993fg}
\bibinfo{author}{\bibfnamefont{H.}~\bibnamefont{Kurki-Suonio}},
  \bibinfo{author}{\bibfnamefont{P.}~\bibnamefont{Laguna}}, \bibnamefont{and}
  \bibinfo{author}{\bibfnamefont{R.~A.} \bibnamefont{Matzner}},
  \bibinfo{journal}{Phys. Rev.} \textbf{\bibinfo{volume}{D48}},
  \bibinfo{pages}{3611} (\bibinfo{year}{1993}), \eprint{astro-ph/9306009}.

\bibitem[{\citenamefont{Deruelle and Goldwirth}(1995)}]{Deruelle:1994pa}
\bibinfo{author}{\bibfnamefont{N.}~\bibnamefont{Deruelle}} \bibnamefont{and}
  \bibinfo{author}{\bibfnamefont{D.~S.} \bibnamefont{Goldwirth}},
  \bibinfo{journal}{Phys. Rev.} \textbf{\bibinfo{volume}{D51}},
  \bibinfo{pages}{1563} (\bibinfo{year}{1995}), \eprint{gr-qc/9409056}.

\bibitem[{\citenamefont{Martin et~al.}(2014{\natexlab{a}})\citenamefont{Martin,
  Ringeval, Trotta, and Vennin}}]{Martin:2013nzq}
\bibinfo{author}{\bibfnamefont{J.}~\bibnamefont{Martin}},
  \bibinfo{author}{\bibfnamefont{C.}~\bibnamefont{Ringeval}},
  \bibinfo{author}{\bibfnamefont{R.}~\bibnamefont{Trotta}}, \bibnamefont{and}
  \bibinfo{author}{\bibfnamefont{V.}~\bibnamefont{Vennin}},
  \bibinfo{journal}{JCAP} \textbf{\bibinfo{volume}{1403}}, \bibinfo{pages}{039}
  (\bibinfo{year}{2014}{\natexlab{a}}), \eprint{1312.3529}.

\bibitem[{\citenamefont{Ijjas et~al.}(2013)\citenamefont{Ijjas, Steinhardt, and
  Loeb}}]{Ijjas_2013}
\bibinfo{author}{\bibfnamefont{A.}~\bibnamefont{Ijjas}},
  \bibinfo{author}{\bibfnamefont{P.~J.} \bibnamefont{Steinhardt}},
  \bibnamefont{and} \bibinfo{author}{\bibfnamefont{A.}~\bibnamefont{Loeb}},
  \bibinfo{journal}{Physics Letters B} \textbf{\bibinfo{volume}{723}},
  \bibinfo{pages}{261–266} (\bibinfo{year}{2013}), ISSN
  \bibinfo{issn}{0370-2693},
  \urlprefix\url{http://dx.doi.org/10.1016/j.physletb.2013.05.023}.

\bibitem[{\citenamefont{Guth et~al.}(2014)\citenamefont{Guth, Kaiser, and
  Nomura}}]{Guth_2014}
\bibinfo{author}{\bibfnamefont{A.~H.} \bibnamefont{Guth}},
  \bibinfo{author}{\bibfnamefont{D.~I.} \bibnamefont{Kaiser}},
  \bibnamefont{and} \bibinfo{author}{\bibfnamefont{Y.}~\bibnamefont{Nomura}},
  \bibinfo{journal}{Physics Letters B} \textbf{\bibinfo{volume}{733}},
  \bibinfo{pages}{112–119} (\bibinfo{year}{2014}), ISSN
  \bibinfo{issn}{0370-2693},
  \urlprefix\url{http://dx.doi.org/10.1016/j.physletb.2014.03.020}.

\bibitem[{\citenamefont{Easther et~al.}(2014)\citenamefont{Easther, Price, and
  Rasero}}]{Easther:2014zga}
\bibinfo{author}{\bibfnamefont{R.}~\bibnamefont{Easther}},
  \bibinfo{author}{\bibfnamefont{L.~C.} \bibnamefont{Price}}, \bibnamefont{and}
  \bibinfo{author}{\bibfnamefont{J.}~\bibnamefont{Rasero}},
  \bibinfo{journal}{JCAP} \textbf{\bibinfo{volume}{1408}}, \bibinfo{pages}{041}
  (\bibinfo{year}{2014}), \eprint{1406.2869}.

\bibitem[{\citenamefont{Ijjas and Steinhardt}(2016)}]{Ijjas_2016}
\bibinfo{author}{\bibfnamefont{A.}~\bibnamefont{Ijjas}} \bibnamefont{and}
  \bibinfo{author}{\bibfnamefont{P.~J.} \bibnamefont{Steinhardt}},
  \bibinfo{journal}{Classical and Quantum Gravity}
  \textbf{\bibinfo{volume}{33}}, \bibinfo{pages}{044001}
  (\bibinfo{year}{2016}), ISSN \bibinfo{issn}{1361-6382},
  \urlprefix\url{http://dx.doi.org/10.1088/0264-9381/33/4/044001}.

\bibitem[{\citenamefont{Chowdhury et~al.}(2019)\citenamefont{Chowdhury, Martin,
  Ringeval, and Vennin}}]{Chowdhury:2019otk}
\bibinfo{author}{\bibfnamefont{D.}~\bibnamefont{Chowdhury}},
  \bibinfo{author}{\bibfnamefont{J.}~\bibnamefont{Martin}},
  \bibinfo{author}{\bibfnamefont{C.}~\bibnamefont{Ringeval}}, \bibnamefont{and}
  \bibinfo{author}{\bibfnamefont{V.}~\bibnamefont{Vennin}},
  \bibinfo{journal}{Phys. Rev. D} \textbf{\bibinfo{volume}{100}},
  \bibinfo{pages}{083537} (\bibinfo{year}{2019}), \eprint{1902.03951}.

\bibitem[{\citenamefont{Brandenberger}(2017)}]{Brandenberger:2016uzh}
\bibinfo{author}{\bibfnamefont{R.}~\bibnamefont{Brandenberger}},
  \bibinfo{journal}{International Journal of Modern Physics D}
  \textbf{\bibinfo{volume}{26}}, \bibinfo{pages}{1740002}
  (\bibinfo{year}{2017}), ISSN \bibinfo{issn}{1793-6594},
  \urlprefix\url{http://dx.doi.org/10.1142/S0218271817400028}.

\bibitem[{\citenamefont{East et~al.}(2016)\citenamefont{East, Kleban, Linde,
  and Senatore}}]{East:2015ggf}
\bibinfo{author}{\bibfnamefont{W.~E.} \bibnamefont{East}},
  \bibinfo{author}{\bibfnamefont{M.}~\bibnamefont{Kleban}},
  \bibinfo{author}{\bibfnamefont{A.}~\bibnamefont{Linde}}, \bibnamefont{and}
  \bibinfo{author}{\bibfnamefont{L.}~\bibnamefont{Senatore}},
  \bibinfo{journal}{Journal of Cosmology and Astroparticle Physics}
  \textbf{\bibinfo{volume}{2016}}, \bibinfo{pages}{010} (\bibinfo{year}{2016}),
  \urlprefix\url{https://doi.org/10.1088%2F1475-7516%2F2016%2F09%2F010}.

\bibitem[{\citenamefont{Clough}(2017)}]{Clough:2017ixw}
\bibinfo{author}{\bibfnamefont{K.}~\bibnamefont{Clough}}, Ph.D. thesis,
  \bibinfo{school}{King's Coll. London}, \bibinfo{address}{Cham}
  (\bibinfo{year}{2017}), \eprint{1704.06811}.

\bibitem[{\citenamefont{Clough et~al.}(2018)\citenamefont{Clough, Flauger, and
  Lim}}]{Clough_2018}
\bibinfo{author}{\bibfnamefont{K.}~\bibnamefont{Clough}},
  \bibinfo{author}{\bibfnamefont{R.}~\bibnamefont{Flauger}}, \bibnamefont{and}
  \bibinfo{author}{\bibfnamefont{E.~A.} \bibnamefont{Lim}},
  \bibinfo{journal}{JCAP} \textbf{\bibinfo{volume}{2018}}, \bibinfo{pages}{065}
  (\bibinfo{year}{2018}), \eprint{1712.07352}.

\bibitem[{\citenamefont{Aurrekoetxea et~al.}(2020)\citenamefont{Aurrekoetxea,
  Clough, Flauger, and Lim}}]{Aurrekoetxea_2020}
\bibinfo{author}{\bibfnamefont{J.~C.} \bibnamefont{Aurrekoetxea}},
  \bibinfo{author}{\bibfnamefont{K.}~\bibnamefont{Clough}},
  \bibinfo{author}{\bibfnamefont{R.}~\bibnamefont{Flauger}}, \bibnamefont{and}
  \bibinfo{author}{\bibfnamefont{E.~A.} \bibnamefont{Lim}},
  \bibinfo{journal}{Journal of Cosmology and Astroparticle Physics}
  \textbf{\bibinfo{volume}{2020}}, \bibinfo{pages}{030–030}
  (\bibinfo{year}{2020}), ISSN \bibinfo{issn}{1475-7516},
  \urlprefix\url{http://dx.doi.org/10.1088/1475-7516/2020/05/030}.

\bibitem[{\citenamefont{Joana and Clesse}(2021)}]{Joana2020}
\bibinfo{author}{\bibfnamefont{C.}~\bibnamefont{Joana}} \bibnamefont{and}
  \bibinfo{author}{\bibfnamefont{S.}~\bibnamefont{Clesse}},
  \bibinfo{journal}{Phys. Rev. D} \textbf{\bibinfo{volume}{103}},
  \bibinfo{pages}{083501} (\bibinfo{year}{2021}),
  \urlprefix\url{https://link.aps.org/doi/10.1103/PhysRevD.103.083501}.

\bibitem[{\citenamefont{Traschen and Brandenberger}(1990)}]{PhysRevD.42.2491}
\bibinfo{author}{\bibfnamefont{J.~H.} \bibnamefont{Traschen}} \bibnamefont{and}
  \bibinfo{author}{\bibfnamefont{R.~H.} \bibnamefont{Brandenberger}},
  \bibinfo{journal}{Phys. Rev. D} \textbf{\bibinfo{volume}{42}},
  \bibinfo{pages}{2491} (\bibinfo{year}{1990}),
  \urlprefix\url{https://link.aps.org/doi/10.1103/PhysRevD.42.2491}.

\bibitem[{\citenamefont{{Dolgov} and {Kirilova}}(1990)}]{1990Dolgov172D}
\bibinfo{author}{\bibfnamefont{A.~D.} \bibnamefont{{Dolgov}}} \bibnamefont{and}
  \bibinfo{author}{\bibfnamefont{D.~P.} \bibnamefont{{Kirilova}}},
  \bibinfo{journal}{Soviet Journal of Nuclear Physics}
  \textbf{\bibinfo{volume}{51}}, \bibinfo{pages}{172} (\bibinfo{year}{1990}).

\bibitem[{\citenamefont{Martin and Ringeval}(2010)}]{Martin2010}
\bibinfo{author}{\bibfnamefont{J.}~\bibnamefont{Martin}} \bibnamefont{and}
  \bibinfo{author}{\bibfnamefont{C.}~\bibnamefont{Ringeval}},
  \bibinfo{journal}{Physical Review D} \textbf{\bibinfo{volume}{82}}
  (\bibinfo{year}{2010}), ISSN \bibinfo{issn}{1550-2368},
  \urlprefix\url{http://dx.doi.org/10.1103/PhysRevD.82.023511}.

\bibitem[{\citenamefont{Martin et~al.}(2015)\citenamefont{Martin, Ringeval, and
  Vennin}}]{Martin2015}
\bibinfo{author}{\bibfnamefont{J.}~\bibnamefont{Martin}},
  \bibinfo{author}{\bibfnamefont{C.}~\bibnamefont{Ringeval}}, \bibnamefont{and}
  \bibinfo{author}{\bibfnamefont{V.}~\bibnamefont{Vennin}},
  \bibinfo{journal}{Physical Review Letters} \textbf{\bibinfo{volume}{114}}
  (\bibinfo{year}{2015}), ISSN \bibinfo{issn}{1079-7114},
  \urlprefix\url{http://dx.doi.org/10.1103/PhysRevLett.114.081303}.

\bibitem[{\citenamefont{Martin et~al.}(2016)\citenamefont{Martin, Ringeval, and
  Vennin}}]{Martin2016}
\bibinfo{author}{\bibfnamefont{J.}~\bibnamefont{Martin}},
  \bibinfo{author}{\bibfnamefont{C.}~\bibnamefont{Ringeval}}, \bibnamefont{and}
  \bibinfo{author}{\bibfnamefont{V.}~\bibnamefont{Vennin}},
  \bibinfo{journal}{Physical Review D} \textbf{\bibinfo{volume}{93}}
  (\bibinfo{year}{2016}), ISSN \bibinfo{issn}{2470-0029},
  \urlprefix\url{http://dx.doi.org/10.1103/PhysRevD.93.103532}.

\bibitem[{\citenamefont{Allahverdi et~al.}(2010)\citenamefont{Allahverdi,
  Brandenberger, Cyr-Racine, and Mazumdar}}]{Brandenberger2010}
\bibinfo{author}{\bibfnamefont{R.}~\bibnamefont{Allahverdi}},
  \bibinfo{author}{\bibfnamefont{R.}~\bibnamefont{Brandenberger}},
  \bibinfo{author}{\bibfnamefont{F.-Y.} \bibnamefont{Cyr-Racine}},
  \bibnamefont{and} \bibinfo{author}{\bibfnamefont{A.}~\bibnamefont{Mazumdar}},
  \bibinfo{journal}{Annual Review of Nuclear and Particle Science}
  \textbf{\bibinfo{volume}{60}}, \bibinfo{pages}{27–51}
  (\bibinfo{year}{2010}), ISSN \bibinfo{issn}{1545-4134},
  \urlprefix\url{http://dx.doi.org/10.1146/annurev.nucl.012809.104511}.

\bibitem[{\citenamefont{Tenkanen and Tomberg}(2020)}]{Tenkanen:2020cvw}
\bibinfo{author}{\bibfnamefont{T.}~\bibnamefont{Tenkanen}} \bibnamefont{and}
  \bibinfo{author}{\bibfnamefont{E.}~\bibnamefont{Tomberg}},
  \bibinfo{journal}{JCAP} \textbf{\bibinfo{volume}{04}}, \bibinfo{pages}{050}
  (\bibinfo{year}{2020}), \eprint{2002.02420}.

\bibitem[{\citenamefont{Kofman et~al.}(1994)\citenamefont{Kofman, Linde, and
  Starobinsky}}]{Kofman1994}
\bibinfo{author}{\bibfnamefont{L.}~\bibnamefont{Kofman}},
  \bibinfo{author}{\bibfnamefont{A.}~\bibnamefont{Linde}}, \bibnamefont{and}
  \bibinfo{author}{\bibfnamefont{A.~A.} \bibnamefont{Starobinsky}},
  \bibinfo{journal}{Physical Review Letters} \textbf{\bibinfo{volume}{73}},
  \bibinfo{pages}{3195–3198} (\bibinfo{year}{1994}), ISSN
  \bibinfo{issn}{0031-9007},
  \urlprefix\url{http://dx.doi.org/10.1103/PhysRevLett.73.3195}.

\bibitem[{\citenamefont{Kofman et~al.}(1997)\citenamefont{Kofman, Linde, and
  Starobinsky}}]{Kofman1997}
\bibinfo{author}{\bibfnamefont{L.}~\bibnamefont{Kofman}},
  \bibinfo{author}{\bibfnamefont{A.}~\bibnamefont{Linde}}, \bibnamefont{and}
  \bibinfo{author}{\bibfnamefont{A.~A.} \bibnamefont{Starobinsky}},
  \bibinfo{journal}{Physical Review D} \textbf{\bibinfo{volume}{56}},
  \bibinfo{pages}{3258–3295} (\bibinfo{year}{1997}), ISSN
  \bibinfo{issn}{1089-4918},
  \urlprefix\url{http://dx.doi.org/10.1103/PhysRevD.56.3258}.

\bibitem[{\citenamefont{Tsujikawa
  et~al.}(1999{\natexlab{a}})\citenamefont{Tsujikawa, Maeda, and
  Torii}}]{Tsujikawa1999A}
\bibinfo{author}{\bibfnamefont{S.}~\bibnamefont{Tsujikawa}},
  \bibinfo{author}{\bibfnamefont{K.-i.} \bibnamefont{Maeda}}, \bibnamefont{and}
  \bibinfo{author}{\bibfnamefont{T.}~\bibnamefont{Torii}},
  \bibinfo{journal}{Phys. Rev. D} \textbf{\bibinfo{volume}{60}},
  \bibinfo{pages}{123505} (\bibinfo{year}{1999}{\natexlab{a}}),
  \urlprefix\url{https://link.aps.org/doi/10.1103/PhysRevD.60.123505}.

\bibitem[{\citenamefont{Tsujikawa
  et~al.}(1999{\natexlab{b}})\citenamefont{Tsujikawa, Maeda, and
  Torii}}]{Tsujikawa1999B}
\bibinfo{author}{\bibfnamefont{S.}~\bibnamefont{Tsujikawa}},
  \bibinfo{author}{\bibfnamefont{K.-i.} \bibnamefont{Maeda}}, \bibnamefont{and}
  \bibinfo{author}{\bibfnamefont{T.}~\bibnamefont{Torii}},
  \bibinfo{journal}{Phys. Rev. D} \textbf{\bibinfo{volume}{60}},
  \bibinfo{pages}{063515} (\bibinfo{year}{1999}{\natexlab{b}}),
  \urlprefix\url{https://link.aps.org/doi/10.1103/PhysRevD.60.063515}.

\bibitem[{\citenamefont{Prokopec and Roos}(1997)}]{Tomislav1997}
\bibinfo{author}{\bibfnamefont{T.}~\bibnamefont{Prokopec}} \bibnamefont{and}
  \bibinfo{author}{\bibfnamefont{T.~G.} \bibnamefont{Roos}},
  \bibinfo{journal}{Physical Review D} \textbf{\bibinfo{volume}{55}},
  \bibinfo{pages}{3768–3775} (\bibinfo{year}{1997}), ISSN
  \bibinfo{issn}{1089-4918},
  \urlprefix\url{http://dx.doi.org/10.1103/PhysRevD.55.3768}.

\bibitem[{\citenamefont{Felder and Kofman}(2001)}]{Gary2001}
\bibinfo{author}{\bibfnamefont{G.}~\bibnamefont{Felder}} \bibnamefont{and}
  \bibinfo{author}{\bibfnamefont{L.}~\bibnamefont{Kofman}},
  \bibinfo{journal}{Physical Review D} \textbf{\bibinfo{volume}{63}}
  (\bibinfo{year}{2001}), ISSN \bibinfo{issn}{1089-4918},
  \urlprefix\url{http://dx.doi.org/10.1103/PhysRevD.63.103503}.

\bibitem[{\citenamefont{García-Bellido
  et~al.}(2003)\citenamefont{García-Bellido, García~Pérez, and
  González-Arroyo}}]{Bellido2003}
\bibinfo{author}{\bibfnamefont{J.}~\bibnamefont{García-Bellido}},
  \bibinfo{author}{\bibfnamefont{M.}~\bibnamefont{García~Pérez}},
  \bibnamefont{and}
  \bibinfo{author}{\bibfnamefont{A.}~\bibnamefont{González-Arroyo}},
  \bibinfo{journal}{Physical Review D} \textbf{\bibinfo{volume}{67}}
  (\bibinfo{year}{2003}), ISSN \bibinfo{issn}{1089-4918},
  \urlprefix\url{http://dx.doi.org/10.1103/PhysRevD.67.103501}.

\bibitem[{\citenamefont{Amin et~al.}(2010)\citenamefont{Amin, Easther, and
  Finkel}}]{Amin2010}
\bibinfo{author}{\bibfnamefont{M.~A.} \bibnamefont{Amin}},
  \bibinfo{author}{\bibfnamefont{R.}~\bibnamefont{Easther}}, \bibnamefont{and}
  \bibinfo{author}{\bibfnamefont{H.}~\bibnamefont{Finkel}},
  \bibinfo{journal}{Journal of Cosmology and Astroparticle Physics}
  \textbf{\bibinfo{volume}{2010}}, \bibinfo{pages}{001–001}
  (\bibinfo{year}{2010}), ISSN \bibinfo{issn}{1475-7516},
  \urlprefix\url{http://dx.doi.org/10.1088/1475-7516/2010/12/001}.

\bibitem[{\citenamefont{Frolov}(2010)}]{Frolov2010}
\bibinfo{author}{\bibfnamefont{A.~V.} \bibnamefont{Frolov}},
  \bibinfo{journal}{Classical and Quantum Gravity}
  \textbf{\bibinfo{volume}{27}}, \bibinfo{pages}{124006}
  (\bibinfo{year}{2010}), ISSN \bibinfo{issn}{1361-6382},
  \urlprefix\url{http://dx.doi.org/10.1088/0264-9381/27/12/124006}.

\bibitem[{\citenamefont{Lozanov and Amin}(2014)}]{Lozanov2014}
\bibinfo{author}{\bibfnamefont{K.~D.} \bibnamefont{Lozanov}} \bibnamefont{and}
  \bibinfo{author}{\bibfnamefont{M.~A.} \bibnamefont{Amin}},
  \bibinfo{journal}{Physical Review D} \textbf{\bibinfo{volume}{90}}
  (\bibinfo{year}{2014}), ISSN \bibinfo{issn}{1550-2368},
  \urlprefix\url{http://dx.doi.org/10.1103/PhysRevD.90.083528}.

\bibitem[{\citenamefont{Repond and Rubio}(2016)}]{Repond2016}
\bibinfo{author}{\bibfnamefont{J.}~\bibnamefont{Repond}} \bibnamefont{and}
  \bibinfo{author}{\bibfnamefont{J.}~\bibnamefont{Rubio}},
  \bibinfo{journal}{Journal of Cosmology and Astroparticle Physics}
  \textbf{\bibinfo{volume}{2016}}, \bibinfo{pages}{043–043}
  (\bibinfo{year}{2016}), ISSN \bibinfo{issn}{1475-7516},
  \urlprefix\url{http://dx.doi.org/10.1088/1475-7516/2016/07/043}.

\bibitem[{\citenamefont{Lozanov and Amin}(2019)}]{Lozanov2019}
\bibinfo{author}{\bibfnamefont{K.~D.} \bibnamefont{Lozanov}} \bibnamefont{and}
  \bibinfo{author}{\bibfnamefont{M.~A.} \bibnamefont{Amin}},
  \bibinfo{journal}{Physical Review D} \textbf{\bibinfo{volume}{99}}
  (\bibinfo{year}{2019}), ISSN \bibinfo{issn}{2470-0029},
  \urlprefix\url{http://dx.doi.org/10.1103/PhysRevD.99.123504}.

\bibitem[{\citenamefont{van~de Bruck et~al.}(2017)\citenamefont{van~de Bruck,
  Dunsby, and Paduraru}}]{Bruck2017}
\bibinfo{author}{\bibfnamefont{C.}~\bibnamefont{van~de Bruck}},
  \bibinfo{author}{\bibfnamefont{P.}~\bibnamefont{Dunsby}}, \bibnamefont{and}
  \bibinfo{author}{\bibfnamefont{L.~E.} \bibnamefont{Paduraru}},
  \bibinfo{journal}{International Journal of Modern Physics D}
  \textbf{\bibinfo{volume}{26}}, \bibinfo{pages}{1750152}
  (\bibinfo{year}{2017}), ISSN \bibinfo{issn}{1793-6594},
  \urlprefix\url{http://dx.doi.org/10.1142/S0218271817501528}.

\bibitem[{\citenamefont{DeCross
  et~al.}(2018{\natexlab{a}})\citenamefont{DeCross, Kaiser, Prabhu,
  Prescod-Weinstein, and Sfakianakis}}]{DeCross2018A}
\bibinfo{author}{\bibfnamefont{M.~P.} \bibnamefont{DeCross}},
  \bibinfo{author}{\bibfnamefont{D.~I.} \bibnamefont{Kaiser}},
  \bibinfo{author}{\bibfnamefont{A.}~\bibnamefont{Prabhu}},
  \bibinfo{author}{\bibfnamefont{C.}~\bibnamefont{Prescod-Weinstein}},
  \bibnamefont{and} \bibinfo{author}{\bibfnamefont{E.~I.}
  \bibnamefont{Sfakianakis}}, \bibinfo{journal}{Physical Review D}
  \textbf{\bibinfo{volume}{97}} (\bibinfo{year}{2018}{\natexlab{a}}), ISSN
  \bibinfo{issn}{2470-0029},
  \urlprefix\url{http://dx.doi.org/10.1103/PhysRevD.97.023526}.

\bibitem[{\citenamefont{DeCross
  et~al.}(2018{\natexlab{b}})\citenamefont{DeCross, Kaiser, Prabhu,
  Prescod-Weinstein, and Sfakianakis}}]{DeCross2018B}
\bibinfo{author}{\bibfnamefont{M.~P.} \bibnamefont{DeCross}},
  \bibinfo{author}{\bibfnamefont{D.~I.} \bibnamefont{Kaiser}},
  \bibinfo{author}{\bibfnamefont{A.}~\bibnamefont{Prabhu}},
  \bibinfo{author}{\bibfnamefont{C.}~\bibnamefont{Prescod-Weinstein}},
  \bibnamefont{and} \bibinfo{author}{\bibfnamefont{E.~I.}
  \bibnamefont{Sfakianakis}}, \bibinfo{journal}{Physical Review D}
  \textbf{\bibinfo{volume}{97}} (\bibinfo{year}{2018}{\natexlab{b}}), ISSN
  \bibinfo{issn}{2470-0029},
  \urlprefix\url{http://dx.doi.org/10.1103/PhysRevD.97.023527}.

\bibitem[{\citenamefont{DeCross
  et~al.}(2018{\natexlab{c}})\citenamefont{DeCross, Kaiser, Prabhu,
  Prescod-Weinstein, and Sfakianakis}}]{DeCross2018C}
\bibinfo{author}{\bibfnamefont{M.~P.} \bibnamefont{DeCross}},
  \bibinfo{author}{\bibfnamefont{D.~I.} \bibnamefont{Kaiser}},
  \bibinfo{author}{\bibfnamefont{A.}~\bibnamefont{Prabhu}},
  \bibinfo{author}{\bibfnamefont{C.}~\bibnamefont{Prescod-Weinstein}},
  \bibnamefont{and} \bibinfo{author}{\bibfnamefont{E.~I.}
  \bibnamefont{Sfakianakis}}, \bibinfo{journal}{Physical Review D}
  \textbf{\bibinfo{volume}{97}} (\bibinfo{year}{2018}{\natexlab{c}}), ISSN
  \bibinfo{issn}{2470-0029},
  \urlprefix\url{http://dx.doi.org/10.1103/PhysRevD.97.023528}.

\bibitem[{\citenamefont{Sfakianakis and van~de Vis}(2019)}]{Sfakianakis2019}
\bibinfo{author}{\bibfnamefont{E.~I.} \bibnamefont{Sfakianakis}}
  \bibnamefont{and} \bibinfo{author}{\bibfnamefont{J.}~\bibnamefont{van~de
  Vis}}, \bibinfo{journal}{Physical Review D} \textbf{\bibinfo{volume}{99}}
  (\bibinfo{year}{2019}), ISSN \bibinfo{issn}{2470-0029},
  \urlprefix\url{http://dx.doi.org/10.1103/PhysRevD.99.083519}.

\bibitem[{\citenamefont{Nguyen et~al.}(2019)\citenamefont{Nguyen, van~de Vis,
  Sfakianakis, Giblin, and Kaiser}}]{Nguyen2019}
\bibinfo{author}{\bibfnamefont{R.}~\bibnamefont{Nguyen}},
  \bibinfo{author}{\bibfnamefont{J.}~\bibnamefont{van~de Vis}},
  \bibinfo{author}{\bibfnamefont{E.~I.} \bibnamefont{Sfakianakis}},
  \bibinfo{author}{\bibfnamefont{J.~T.} \bibnamefont{Giblin}},
  \bibnamefont{and} \bibinfo{author}{\bibfnamefont{D.~I.}
  \bibnamefont{Kaiser}}, \bibinfo{journal}{Physical Review Letters}
  \textbf{\bibinfo{volume}{123}} (\bibinfo{year}{2019}), ISSN
  \bibinfo{issn}{1079-7114},
  \urlprefix\url{http://dx.doi.org/10.1103/PhysRevLett.123.171301}.

\bibitem[{\citenamefont{Rubio and Tomberg}(2019)}]{Rubio2019}
\bibinfo{author}{\bibfnamefont{J.}~\bibnamefont{Rubio}} \bibnamefont{and}
  \bibinfo{author}{\bibfnamefont{E.~S.} \bibnamefont{Tomberg}},
  \bibinfo{journal}{Journal of Cosmology and Astroparticle Physics}
  \textbf{\bibinfo{volume}{2019}}, \bibinfo{pages}{021–021}
  (\bibinfo{year}{2019}), ISSN \bibinfo{issn}{1475-7516},
  \urlprefix\url{http://dx.doi.org/10.1088/1475-7516/2019/04/021}.

\bibitem[{\citenamefont{van~de Vis et~al.}(2020)\citenamefont{van~de Vis,
  Nguyen, Sfakianakis, Giblin, and Kaiser}}]{Vis2020}
\bibinfo{author}{\bibfnamefont{J.}~\bibnamefont{van~de Vis}},
  \bibinfo{author}{\bibfnamefont{R.}~\bibnamefont{Nguyen}},
  \bibinfo{author}{\bibfnamefont{E.~I.} \bibnamefont{Sfakianakis}},
  \bibinfo{author}{\bibfnamefont{J.~T.} \bibnamefont{Giblin}},
  \bibnamefont{and} \bibinfo{author}{\bibfnamefont{D.~I.}
  \bibnamefont{Kaiser}}, \bibinfo{journal}{Physical Review D}
  \textbf{\bibinfo{volume}{102}} (\bibinfo{year}{2020}), ISSN
  \bibinfo{issn}{2470-0029},
  \urlprefix\url{http://dx.doi.org/10.1103/PhysRevD.102.043528}.

\bibitem[{\citenamefont{Ema et~al.}(2017)\citenamefont{Ema, Jinno, Mukaida, and
  Nakayama}}]{Ema2017}
\bibinfo{author}{\bibfnamefont{Y.}~\bibnamefont{Ema}},
  \bibinfo{author}{\bibfnamefont{R.}~\bibnamefont{Jinno}},
  \bibinfo{author}{\bibfnamefont{K.}~\bibnamefont{Mukaida}}, \bibnamefont{and}
  \bibinfo{author}{\bibfnamefont{K.}~\bibnamefont{Nakayama}},
  \bibinfo{journal}{Journal of Cosmology and Astroparticle Physics}
  \textbf{\bibinfo{volume}{2017}}, \bibinfo{pages}{045–045}
  (\bibinfo{year}{2017}), ISSN \bibinfo{issn}{1475-7516},
  \urlprefix\url{http://dx.doi.org/10.1088/1475-7516/2017/02/045}.

\bibitem[{\citenamefont{Ema et~al.}(2021)\citenamefont{Ema, Jinno, Nakayama,
  and van~de Vis}}]{Ema2021}
\bibinfo{author}{\bibfnamefont{Y.}~\bibnamefont{Ema}},
  \bibinfo{author}{\bibfnamefont{R.}~\bibnamefont{Jinno}},
  \bibinfo{author}{\bibfnamefont{K.}~\bibnamefont{Nakayama}}, \bibnamefont{and}
  \bibinfo{author}{\bibfnamefont{J.}~\bibnamefont{van~de Vis}},
  \bibinfo{journal}{Physical Review D} \textbf{\bibinfo{volume}{103}}
  (\bibinfo{year}{2021}), ISSN \bibinfo{issn}{2470-0029},
  \urlprefix\url{http://dx.doi.org/10.1103/PhysRevD.103.103536}.

\bibitem[{\citenamefont{Hamada et~al.}(2021)\citenamefont{Hamada, Kawana, and
  Scherlis}}]{Hamada2021}
\bibinfo{author}{\bibfnamefont{Y.}~\bibnamefont{Hamada}},
  \bibinfo{author}{\bibfnamefont{K.}~\bibnamefont{Kawana}}, \bibnamefont{and}
  \bibinfo{author}{\bibfnamefont{A.}~\bibnamefont{Scherlis}},
  \bibinfo{journal}{Journal of Cosmology and Astroparticle Physics}
  \textbf{\bibinfo{volume}{2021}}, \bibinfo{pages}{062} (\bibinfo{year}{2021}),
  ISSN \bibinfo{issn}{1475-7516},
  \urlprefix\url{http://dx.doi.org/10.1088/1475-7516/2021/03/062}.

\bibitem[{\citenamefont{Bassett et~al.}(1999)\citenamefont{Bassett, Kaiser, and
  Maartens}}]{Bassett1999}
\bibinfo{author}{\bibfnamefont{B.~A.} \bibnamefont{Bassett}},
  \bibinfo{author}{\bibfnamefont{D.~I.} \bibnamefont{Kaiser}},
  \bibnamefont{and} \bibinfo{author}{\bibfnamefont{R.}~\bibnamefont{Maartens}},
  \bibinfo{journal}{Physics Letters B} \textbf{\bibinfo{volume}{455}},
  \bibinfo{pages}{84–89} (\bibinfo{year}{1999}), ISSN
  \bibinfo{issn}{0370-2693},
  \urlprefix\url{http://dx.doi.org/10.1016/S0370-2693(99)00478-5}.

\bibitem[{\citenamefont{Jedamzik
  et~al.}(2010{\natexlab{a}})\citenamefont{Jedamzik, Lemoine, and
  Martin}}]{Jedamzik2010A}
\bibinfo{author}{\bibfnamefont{K.}~\bibnamefont{Jedamzik}},
  \bibinfo{author}{\bibfnamefont{M.}~\bibnamefont{Lemoine}}, \bibnamefont{and}
  \bibinfo{author}{\bibfnamefont{J.}~\bibnamefont{Martin}},
  \bibinfo{journal}{Journal of Cosmology and Astroparticle Physics}
  \textbf{\bibinfo{volume}{2010}}, \bibinfo{pages}{034–034}
  (\bibinfo{year}{2010}{\natexlab{a}}), ISSN \bibinfo{issn}{1475-7516},
  \urlprefix\url{http://dx.doi.org/10.1088/1475-7516/2010/09/034}.

\bibitem[{\citenamefont{Jedamzik
  et~al.}(2010{\natexlab{b}})\citenamefont{Jedamzik, Lemoine, and
  Martin}}]{Jedamzik2010B}
\bibinfo{author}{\bibfnamefont{K.}~\bibnamefont{Jedamzik}},
  \bibinfo{author}{\bibfnamefont{M.}~\bibnamefont{Lemoine}}, \bibnamefont{and}
  \bibinfo{author}{\bibfnamefont{J.}~\bibnamefont{Martin}},
  \bibinfo{journal}{Journal of Cosmology and Astroparticle Physics}
  \textbf{\bibinfo{volume}{2010}}, \bibinfo{pages}{021–021}
  (\bibinfo{year}{2010}{\natexlab{b}}), ISSN \bibinfo{issn}{1475-7516},
  \urlprefix\url{http://dx.doi.org/10.1088/1475-7516/2010/04/021}.

\bibitem[{\citenamefont{Zhou et~al.}(2020)\citenamefont{Zhou, Jiang, Cai,
  Sasaki, and Pi}}]{Zihan2020}
\bibinfo{author}{\bibfnamefont{Z.}~\bibnamefont{Zhou}},
  \bibinfo{author}{\bibfnamefont{J.}~\bibnamefont{Jiang}},
  \bibinfo{author}{\bibfnamefont{Y.-F.} \bibnamefont{Cai}},
  \bibinfo{author}{\bibfnamefont{M.}~\bibnamefont{Sasaki}}, \bibnamefont{and}
  \bibinfo{author}{\bibfnamefont{S.}~\bibnamefont{Pi}},
  \bibinfo{journal}{Physical Review D} \textbf{\bibinfo{volume}{102}}
  (\bibinfo{year}{2020}), ISSN \bibinfo{issn}{2470-0029},
  \urlprefix\url{http://dx.doi.org/10.1103/PhysRevD.102.103527}.

\bibitem[{\citenamefont{Giblin and Tishue}(2019)}]{Giblin2019}
\bibinfo{author}{\bibfnamefont{J.~T.} \bibnamefont{Giblin}} \bibnamefont{and}
  \bibinfo{author}{\bibfnamefont{A.~J.} \bibnamefont{Tishue}},
  \bibinfo{journal}{Phys. Rev. D} \textbf{\bibinfo{volume}{100}},
  \bibinfo{pages}{063543} (\bibinfo{year}{2019}),
  \urlprefix\url{https://link.aps.org/doi/10.1103/PhysRevD.100.063543}.

\bibitem[{\citenamefont{Kou et~al.}(2021{\natexlab{a}})\citenamefont{Kou, Tian,
  and Zhou}}]{Kou:2019bbc}
\bibinfo{author}{\bibfnamefont{X.-X.} \bibnamefont{Kou}},
  \bibinfo{author}{\bibfnamefont{C.}~\bibnamefont{Tian}}, \bibnamefont{and}
  \bibinfo{author}{\bibfnamefont{S.-Y.} \bibnamefont{Zhou}},
  \bibinfo{journal}{Class. Quant. Grav.} \textbf{\bibinfo{volume}{38}},
  \bibinfo{pages}{045005} (\bibinfo{year}{2021}{\natexlab{a}}),
  \eprint{1912.09658}.

\bibitem[{\citenamefont{Kou et~al.}(2021{\natexlab{b}})\citenamefont{Kou,
  Mertens, Tian, and Zhou}}]{Kou:2021bij}
\bibinfo{author}{\bibfnamefont{X.-X.} \bibnamefont{Kou}},
  \bibinfo{author}{\bibfnamefont{J.~B.} \bibnamefont{Mertens}},
  \bibinfo{author}{\bibfnamefont{C.}~\bibnamefont{Tian}}, \bibnamefont{and}
  \bibinfo{author}{\bibfnamefont{S.-Y.} \bibnamefont{Zhou}}
  (\bibinfo{year}{2021}{\natexlab{b}}), \eprint{2112.07626}.

\bibitem[{\citenamefont{Bezrukov and Shaposhnikov}(2008)}]{Bezrukov:2007ep}
\bibinfo{author}{\bibfnamefont{F.~L.} \bibnamefont{Bezrukov}} \bibnamefont{and}
  \bibinfo{author}{\bibfnamefont{M.}~\bibnamefont{Shaposhnikov}},
  \bibinfo{journal}{Phys. Lett. B} \textbf{\bibinfo{volume}{659}},
  \bibinfo{pages}{703} (\bibinfo{year}{2008}), \eprint{0710.3755}.

\bibitem[{\citenamefont{García-Bellido
  et~al.}(2009)\citenamefont{García-Bellido, Figueroa, and
  Rubio}}]{Bellido2009}
\bibinfo{author}{\bibfnamefont{J.}~\bibnamefont{García-Bellido}},
  \bibinfo{author}{\bibfnamefont{D.~G.} \bibnamefont{Figueroa}},
  \bibnamefont{and} \bibinfo{author}{\bibfnamefont{J.}~\bibnamefont{Rubio}},
  \bibinfo{journal}{Physical Review D} \textbf{\bibinfo{volume}{79}}
  (\bibinfo{year}{2009}), ISSN \bibinfo{issn}{1550-2368},
  \urlprefix\url{http://dx.doi.org/10.1103/PhysRevD.79.063531}.

\bibitem[{\citenamefont{Callan et~al.}(1970)\citenamefont{Callan, Coleman, and
  Jackiw}}]{1970Callan}
\bibinfo{author}{\bibfnamefont{J.}~\bibnamefont{Callan},
  \bibfnamefont{Curtis~G.}},
  \bibinfo{author}{\bibfnamefont{S.}~\bibnamefont{Coleman}}, \bibnamefont{and}
  \bibinfo{author}{\bibfnamefont{R.}~\bibnamefont{Jackiw}},
  \bibinfo{journal}{Annals of Physics} \textbf{\bibinfo{volume}{59}},
  \bibinfo{pages}{42} (\bibinfo{year}{1970}).

\bibitem[{\citenamefont{Martin et~al.}(2014{\natexlab{b}})\citenamefont{Martin,
  Ringeval, and Vennin}}]{MARTIN201475}
\bibinfo{author}{\bibfnamefont{J.}~\bibnamefont{Martin}},
  \bibinfo{author}{\bibfnamefont{C.}~\bibnamefont{Ringeval}}, \bibnamefont{and}
  \bibinfo{author}{\bibfnamefont{V.}~\bibnamefont{Vennin}},
  \bibinfo{journal}{Physics of the Dark Universe}
  \textbf{\bibinfo{volume}{5-6}}, \bibinfo{pages}{75}
  (\bibinfo{year}{2014}{\natexlab{b}}), ISSN \bibinfo{issn}{2212-6864},
  \urlprefix\url{https://www.sciencedirect.com/science/article/pii/S2212686414000053}.

\bibitem[{\citenamefont{Degrassi et~al.}(2012)\citenamefont{Degrassi, Vita,
  Elias-Mir{\'{o}}, Espinosa, Giudice, Isidori, and Strumia}}]{Degrassi_2012}
\bibinfo{author}{\bibfnamefont{G.}~\bibnamefont{Degrassi}},
  \bibinfo{author}{\bibfnamefont{S.~D.} \bibnamefont{Vita}},
  \bibinfo{author}{\bibfnamefont{J.}~\bibnamefont{Elias-Mir{\'{o}}}},
  \bibinfo{author}{\bibfnamefont{J.~R.} \bibnamefont{Espinosa}},
  \bibinfo{author}{\bibfnamefont{G.~F.} \bibnamefont{Giudice}},
  \bibinfo{author}{\bibfnamefont{G.}~\bibnamefont{Isidori}}, \bibnamefont{and}
  \bibinfo{author}{\bibfnamefont{A.}~\bibnamefont{Strumia}},
  \bibinfo{journal}{Journal of High Energy Physics}
  \textbf{\bibinfo{volume}{2012}} (\bibinfo{year}{2012}),
  \urlprefix\url{https://doi.org/10.1007%2Fjhep08%282012%29098}.

\bibitem[{\citenamefont{Tanabashi et~al.}(2018)\citenamefont{Tanabashi,
  Hagiwara, Hikasa, Nakamura, Sumino, Takahashi, Tanaka, Agashe, Aielli, Amsler
  et~al.}}]{PhysRevD.98.030001}
\bibinfo{author}{\bibfnamefont{M.}~\bibnamefont{Tanabashi}},
  \bibinfo{author}{\bibfnamefont{K.}~\bibnamefont{Hagiwara}},
  \bibinfo{author}{\bibfnamefont{K.}~\bibnamefont{Hikasa}},
  \bibinfo{author}{\bibfnamefont{K.}~\bibnamefont{Nakamura}},
  \bibinfo{author}{\bibfnamefont{Y.}~\bibnamefont{Sumino}},
  \bibinfo{author}{\bibfnamefont{F.}~\bibnamefont{Takahashi}},
  \bibinfo{author}{\bibfnamefont{J.}~\bibnamefont{Tanaka}},
  \bibinfo{author}{\bibfnamefont{K.}~\bibnamefont{Agashe}},
  \bibinfo{author}{\bibfnamefont{G.}~\bibnamefont{Aielli}},
  \bibinfo{author}{\bibfnamefont{C.}~\bibnamefont{Amsler}},
  \bibnamefont{et~al.} (\bibinfo{collaboration}{Particle Data Group}),
  \bibinfo{journal}{Phys. Rev. D} \textbf{\bibinfo{volume}{98}},
  \bibinfo{pages}{030001} (\bibinfo{year}{2018}),
  \urlprefix\url{https://link.aps.org/doi/10.1103/PhysRevD.98.030001}.

\bibitem[{\citenamefont{Branchina et~al.}(2019)\citenamefont{Branchina,
  Bentivegna, Contino, and Zappalà}}]{Branchina2019}
\bibinfo{author}{\bibfnamefont{V.}~\bibnamefont{Branchina}},
  \bibinfo{author}{\bibfnamefont{E.}~\bibnamefont{Bentivegna}},
  \bibinfo{author}{\bibfnamefont{F.}~\bibnamefont{Contino}}, \bibnamefont{and}
  \bibinfo{author}{\bibfnamefont{D.}~\bibnamefont{Zappalà}},
  \bibinfo{journal}{Physical Review D} \textbf{\bibinfo{volume}{99}}
  (\bibinfo{year}{2019}), ISSN \bibinfo{issn}{2470-0029},
  \urlprefix\url{http://dx.doi.org/10.1103/PhysRevD.99.096029}.

\bibitem[{\citenamefont{Linde}(1984)}]{Linde:1983mx}
\bibinfo{author}{\bibfnamefont{A.~D.} \bibnamefont{Linde}},
  \bibinfo{journal}{Lett. Nuovo Cim.} \textbf{\bibinfo{volume}{39}},
  \bibinfo{pages}{401} (\bibinfo{year}{1984}).

\bibitem[{\citenamefont{Linde}(2018)}]{Linde:2017pwt}
\bibinfo{author}{\bibfnamefont{A.}~\bibnamefont{Linde}},
  \bibinfo{journal}{Found. Phys.} \textbf{\bibinfo{volume}{48}},
  \bibinfo{pages}{1246} (\bibinfo{year}{2018}), \eprint{1710.04278}.

\bibitem[{\citenamefont{Clough et~al.}(2015)\citenamefont{Clough, Figueras,
  Finkel, Kunesch, Lim, and Tunyasuvunakool}}]{Clough_2015}
\bibinfo{author}{\bibfnamefont{K.}~\bibnamefont{Clough}},
  \bibinfo{author}{\bibfnamefont{P.}~\bibnamefont{Figueras}},
  \bibinfo{author}{\bibfnamefont{H.}~\bibnamefont{Finkel}},
  \bibinfo{author}{\bibfnamefont{M.}~\bibnamefont{Kunesch}},
  \bibinfo{author}{\bibfnamefont{E.~A.} \bibnamefont{Lim}}, \bibnamefont{and}
  \bibinfo{author}{\bibfnamefont{S.}~\bibnamefont{Tunyasuvunakool}},
  \bibinfo{journal}{Classical and Quantum Gravity}
  \textbf{\bibinfo{volume}{32}}, \bibinfo{pages}{245011}
  (\bibinfo{year}{2015}), ISSN \bibinfo{issn}{1361-6382}, \eprint{1503.03436},
  \urlprefix\url{http://dx.doi.org/10.1088/0264-9381/32/24/245011}.

\bibitem[{\citenamefont{Andrade et~al.}(2021)\citenamefont{Andrade, Salo,
  Aurrekoetxea, Bamber, Clough, Croft, de~Jong, Drew, Duran, Ferreira
  et~al.}}]{Andrade2021}
\bibinfo{author}{\bibfnamefont{T.}~\bibnamefont{Andrade}},
  \bibinfo{author}{\bibfnamefont{L.~A.} \bibnamefont{Salo}},
  \bibinfo{author}{\bibfnamefont{J.~C.} \bibnamefont{Aurrekoetxea}},
  \bibinfo{author}{\bibfnamefont{J.}~\bibnamefont{Bamber}},
  \bibinfo{author}{\bibfnamefont{K.}~\bibnamefont{Clough}},
  \bibinfo{author}{\bibfnamefont{R.}~\bibnamefont{Croft}},
  \bibinfo{author}{\bibfnamefont{E.}~\bibnamefont{de~Jong}},
  \bibinfo{author}{\bibfnamefont{A.}~\bibnamefont{Drew}},
  \bibinfo{author}{\bibfnamefont{A.}~\bibnamefont{Duran}},
  \bibinfo{author}{\bibfnamefont{P.~G.} \bibnamefont{Ferreira}},
  \bibnamefont{et~al.}, \bibinfo{journal}{Journal of Open Source Software}
  \textbf{\bibinfo{volume}{6}}, \bibinfo{pages}{3703} (\bibinfo{year}{2021}),
  \urlprefix\url{https://doi.org/10.21105/joss.03703}.

\bibitem[{\citenamefont{Gourgoulhon}(2007)}]{Gourgoulhon:2007ue}
\bibinfo{author}{\bibfnamefont{E.}~\bibnamefont{Gourgoulhon}}
  (\bibinfo{year}{2007}), \eprint{gr-qc/0703035}.

\bibitem[{\citenamefont{Garfinkle and Mead}(2020)}]{Garfinkle_2020}
\bibinfo{author}{\bibfnamefont{D.}~\bibnamefont{Garfinkle}} \bibnamefont{and}
  \bibinfo{author}{\bibfnamefont{L.}~\bibnamefont{Mead}},
  \bibinfo{journal}{Physical Review D} \textbf{\bibinfo{volume}{102}}
  (\bibinfo{year}{2020}), ISSN \bibinfo{issn}{2470-0029},
  \urlprefix\url{http://dx.doi.org/10.1103/PhysRevD.102.044022}.

\bibitem[{\citenamefont{Shibata and Nakamura}(1995)}]{PhysRevD.52.5428}
\bibinfo{author}{\bibfnamefont{M.}~\bibnamefont{Shibata}} \bibnamefont{and}
  \bibinfo{author}{\bibfnamefont{T.}~\bibnamefont{Nakamura}},
  \bibinfo{journal}{Phys. Rev. D} \textbf{\bibinfo{volume}{52}},
  \bibinfo{pages}{5428} (\bibinfo{year}{1995}),
  \urlprefix\url{https://link.aps.org/doi/10.1103/PhysRevD.52.5428}.

\bibitem[{\citenamefont{Baumgarte and Shapiro}(1998)}]{Baumgarte_1998}
\bibinfo{author}{\bibfnamefont{T.~W.} \bibnamefont{Baumgarte}}
  \bibnamefont{and} \bibinfo{author}{\bibfnamefont{S.~L.}
  \bibnamefont{Shapiro}}, \bibinfo{journal}{Physical Review D}
  \textbf{\bibinfo{volume}{59}} (\bibinfo{year}{1998}), ISSN
  \bibinfo{issn}{1089-4918},
  \urlprefix\url{http://dx.doi.org/10.1103/PhysRevD.59.024007}.

\bibitem[{\citenamefont{Nakamura et~al.}(1987)\citenamefont{Nakamura, Oohara,
  and Kojima}}]{10.1143/PTPS.90.1}
\bibinfo{author}{\bibfnamefont{T.}~\bibnamefont{Nakamura}},
  \bibinfo{author}{\bibfnamefont{K.}~\bibnamefont{Oohara}}, \bibnamefont{and}
  \bibinfo{author}{\bibfnamefont{Y.}~\bibnamefont{Kojima}},
  \bibinfo{journal}{Progress of Theoretical Physics Supplement}
  \textbf{\bibinfo{volume}{90}}, \bibinfo{pages}{1} (\bibinfo{year}{1987}),
  ISSN \bibinfo{issn}{0375-9687},
  \urlprefix\url{https://doi.org/10.1143/PTPS.90.1}.

\bibitem[{\citenamefont{Nazari et~al.}(2021)\citenamefont{Nazari, Cicoli,
  Clough, and Muia}}]{Nazari2021}
\bibinfo{author}{\bibfnamefont{Z.}~\bibnamefont{Nazari}},
  \bibinfo{author}{\bibfnamefont{M.}~\bibnamefont{Cicoli}},
  \bibinfo{author}{\bibfnamefont{K.}~\bibnamefont{Clough}}, \bibnamefont{and}
  \bibinfo{author}{\bibfnamefont{F.}~\bibnamefont{Muia}},
  \bibinfo{journal}{Journal of Cosmology and Astroparticle Physics}
  \textbf{\bibinfo{volume}{2021}}, \bibinfo{pages}{027} (\bibinfo{year}{2021}),
  ISSN \bibinfo{issn}{1475-7516},
  \urlprefix\url{http://dx.doi.org/10.1088/1475-7516/2021/05/027}.

\bibitem[{\citenamefont{de~Jong et~al.}(2022)\citenamefont{de~Jong,
  Aurrekoetxea, and Lim}}]{de_Jong_2022}
\bibinfo{author}{\bibfnamefont{E.}~\bibnamefont{de~Jong}},
  \bibinfo{author}{\bibfnamefont{J.~C.} \bibnamefont{Aurrekoetxea}},
  \bibnamefont{and} \bibinfo{author}{\bibfnamefont{E.~A.} \bibnamefont{Lim}},
  \bibinfo{journal}{Journal of Cosmology and Astroparticle Physics}
  \textbf{\bibinfo{volume}{2022}}, \bibinfo{pages}{029} (\bibinfo{year}{2022}),
  \urlprefix\url{https://doi.org/10.1088%2F1475-7516%2F2022%2F03%2F029}.

\bibitem[{\citenamefont{Es-haghi and Sheykhi}(2020)}]{Es-haghi:2020oab}
\bibinfo{author}{\bibfnamefont{M.}~\bibnamefont{Es-haghi}} \bibnamefont{and}
  \bibinfo{author}{\bibfnamefont{A.}~\bibnamefont{Sheykhi}}
  (\bibinfo{year}{2020}), \eprint{2012.08035}.

\bibitem[{\citenamefont{Lee et~al.}(2022)\citenamefont{Lee, Modak, Oda, and
  Takahashi}}]{Lee:2021rzy}
\bibinfo{author}{\bibfnamefont{S.~M.} \bibnamefont{Lee}},
  \bibinfo{author}{\bibfnamefont{T.}~\bibnamefont{Modak}},
  \bibinfo{author}{\bibfnamefont{K.-y.} \bibnamefont{Oda}}, \bibnamefont{and}
  \bibinfo{author}{\bibfnamefont{T.}~\bibnamefont{Takahashi}},
  \bibinfo{journal}{Eur. Phys. J. C} \textbf{\bibinfo{volume}{82}},
  \bibinfo{pages}{18} (\bibinfo{year}{2022}), \eprint{2108.02383}.

\bibitem[{\citenamefont{Lin and Sasaki}(2016)}]{Chunshan2016}
\bibinfo{author}{\bibfnamefont{C.}~\bibnamefont{Lin}} \bibnamefont{and}
  \bibinfo{author}{\bibfnamefont{M.}~\bibnamefont{Sasaki}},
  \bibinfo{journal}{Physics Letters B} \textbf{\bibinfo{volume}{752}},
  \bibinfo{pages}{84–88} (\bibinfo{year}{2016}), ISSN
  \bibinfo{issn}{0370-2693},
  \urlprefix\url{http://dx.doi.org/10.1016/j.physletb.2015.11.021}.

\bibitem[{\citenamefont{Cai et~al.}(2021)\citenamefont{Cai, Jiang, Sasaki,
  Vardanyan, and Zhou}}]{Cai2021}
\bibinfo{author}{\bibfnamefont{Y.-F.} \bibnamefont{Cai}},
  \bibinfo{author}{\bibfnamefont{J.}~\bibnamefont{Jiang}},
  \bibinfo{author}{\bibfnamefont{M.}~\bibnamefont{Sasaki}},
  \bibinfo{author}{\bibfnamefont{V.}~\bibnamefont{Vardanyan}},
  \bibnamefont{and} \bibinfo{author}{\bibfnamefont{Z.}~\bibnamefont{Zhou}},
  \bibinfo{journal}{Phys. Rev. Lett.} \textbf{\bibinfo{volume}{127}},
  \bibinfo{pages}{251301} (\bibinfo{year}{2021}),
  \urlprefix\url{https://link.aps.org/doi/10.1103/PhysRevLett.127.251301}.

\bibitem[{\citenamefont{Bethel et~al.}(2012)\citenamefont{Bethel, Childs, and
  Hansen}}]{10.5555/2422936}
\bibinfo{author}{\bibfnamefont{E.~W.} \bibnamefont{Bethel}},
  \bibinfo{author}{\bibfnamefont{H.}~\bibnamefont{Childs}}, \bibnamefont{and}
  \bibinfo{author}{\bibfnamefont{C.}~\bibnamefont{Hansen}},
  \emph{\bibinfo{title}{High Performance Visualization: Enabling Extreme-Scale
  Scientific Insight}} (\bibinfo{publisher}{Chapman; Hall/CRC},
  \bibinfo{year}{2012}), \bibinfo{edition}{1st} ed., ISBN
  \bibinfo{isbn}{1439875723}.

\bibitem[{\citenamefont{Turk et~al.}(2010)\citenamefont{Turk, Smith, Oishi,
  Skory, Skillman, Abel, and Norman}}]{Turk_2010}
\bibinfo{author}{\bibfnamefont{M.~J.} \bibnamefont{Turk}},
  \bibinfo{author}{\bibfnamefont{B.~D.} \bibnamefont{Smith}},
  \bibinfo{author}{\bibfnamefont{J.~S.} \bibnamefont{Oishi}},
  \bibinfo{author}{\bibfnamefont{S.}~\bibnamefont{Skory}},
  \bibinfo{author}{\bibfnamefont{S.~W.} \bibnamefont{Skillman}},
  \bibinfo{author}{\bibfnamefont{T.}~\bibnamefont{Abel}}, \bibnamefont{and}
  \bibinfo{author}{\bibfnamefont{M.~L.} \bibnamefont{Norman}},
  \bibinfo{journal}{The Astrophysical Journal Supplement Series}
  \textbf{\bibinfo{volume}{192}}, \bibinfo{pages}{9} (\bibinfo{year}{2010}),
  ISSN \bibinfo{issn}{1538-4365},
  \urlprefix\url{http://dx.doi.org/10.1088/0067-0049/192/1/9}.

\bibitem[{\citenamefont{Baker et~al.}(2006)\citenamefont{Baker, Centrella,
  Choi, Koppitz, and van Meter}}]{Baker_2006}
\bibinfo{author}{\bibfnamefont{J.~G.} \bibnamefont{Baker}},
  \bibinfo{author}{\bibfnamefont{J.}~\bibnamefont{Centrella}},
  \bibinfo{author}{\bibfnamefont{D.-I.} \bibnamefont{Choi}},
  \bibinfo{author}{\bibfnamefont{M.}~\bibnamefont{Koppitz}}, \bibnamefont{and}
  \bibinfo{author}{\bibfnamefont{J.}~\bibnamefont{van Meter}},
  \bibinfo{journal}{Physical Review Letters} \textbf{\bibinfo{volume}{96}}
  (\bibinfo{year}{2006}), ISSN \bibinfo{issn}{1079-7114},
  \urlprefix\url{http://dx.doi.org/10.1103/PhysRevLett.96.111102}.

\bibitem[{\citenamefont{Campanelli et~al.}(2006)\citenamefont{Campanelli,
  Lousto, Marronetti, and Zlochower}}]{Campanelli_2006}
\bibinfo{author}{\bibfnamefont{M.}~\bibnamefont{Campanelli}},
  \bibinfo{author}{\bibfnamefont{C.~O.} \bibnamefont{Lousto}},
  \bibinfo{author}{\bibfnamefont{P.}~\bibnamefont{Marronetti}},
  \bibnamefont{and}
  \bibinfo{author}{\bibfnamefont{Y.}~\bibnamefont{Zlochower}},
  \bibinfo{journal}{Physical Review Letters} \textbf{\bibinfo{volume}{96}}
  (\bibinfo{year}{2006}), ISSN \bibinfo{issn}{1079-7114},
  \urlprefix\url{http://dx.doi.org/10.1103/PhysRevLett.96.111101}.

\end{thebibliography}


\appendix

\section{BSSN formalism of numerical relativity}\label{sec:BSNN}

In this work, we solve the BSSN formulation of the Einstein equations using \texttt{GRChombo}~\cite{Clough_2015,Andrade2021}, a multipurpose numerical relativity code. In the context of the 3+1 decomposition of General Relativity, the line element reads
\be\label{timeline_AP}
\rr d s^2 = - \alpha^2 \rr d t^2 + \gamma_{ij}(\rr d x^i + \beta^i \rr d t)(\rr d x^j + \beta^j \rr d t)
\ee
where $\gamma\ij$ is the metric of the 3-dimensional hypersurface, and the lapse and shift gauge parameters are given by $\alpha(t)$ and $\beta^i(t)$ {respectively}. A further conformal decomposition of the 3-metric follows,
\be
\gm\ij = \frac1\chi \tgm\ij = \psi^4\tgm\ij \quad \text{with } \text{ det}(\tgm\ij) = 1 ~, 
\ee
where $\chi$ and $\psi$ are two different parametrisations of the metric conformal factor. While the former is used during the temporal integration, the latter is preferred when constructing the initial conditions. The extrinsic curvature is thus split in $\tA\ij$ and $K$, respectively, the conformal traceless part and its trace, 
\be
K\ij = \frac1\chi \left( \tA\ij +\frac13\tgm\ij K\right)~.
\ee
In addition, the first spatial derivatives of the metric are considered as dynamical variables
\be
\tilde\Gamma^i \equiv \tgm^{jk} \tilde\Gamma^i_{jk} = - \partial_j\tgm\ij ~,
\ee
where $ \tilde\Gamma^i_{jk} $ are the Christoffel symbols associated with the conformal metric $ \tilde\gamma_{ij} $.

\subsection{Evolution equations}

The evolution equations for the BSSN variables are then given by 
\begin{align} 
&\partial_t\chi=\frac{2}{3}\,\alpha\,\chi\, K - \frac{2}{3}\,\chi \,\partial_k \beta^k + \beta^k\,\partial_k \chi ~ , \label{eqn:dtchi2} \\
&\partial_t\tilde\gamma_{ij} =-2\,\alpha\, \tA_{ij}+\tilde \gamma_{ik}\,\partial_j\beta^k+\tilde \gamma_{jk}\,\partial_i\beta^k \nonumber \\
&\hspace{1.3cm} -\frac{2}{3}\,\tilde \gamma_{ij}\,\partial_k\beta^k +\beta^k\,\partial_k \tilde \gamma_{ij} ~ , \label{eqn:dttgamma2} \\
&\partial_t K = -\gamma^{ij}D_i D_j \alpha + \alpha\left(\tilde{A}_{ij} \tilde{A}^{ij} + \frac{1}{3} K^2 \right) \nonumber \\
&\hspace{1.3cm} + \beta^i\partial_iK + 4\pi\,\alpha(\rho_{\rm sf}+ S) \label{eqn:dtK2} ~ , 
 \end{align}
 \begin{align} 
&\partial_t\tA_{ij} = \left[- \chi D_iD_j \alpha + \chi \alpha\left( R_{ij} - 8\pi\, \,S_{ij}\right)\right]^\textrm{TF} \nonumber \\
&\hspace{1.3cm} + \alpha (K \tA_{ij} - 2 \tA_{il}\,\tA^l{}_j) \nonumber \\
&\hspace{1.3cm} + \tA_{ik}\, \partial_j\beta^k + \tA_{jk}\,\partial_i\beta^k \nonumber \\
&\hspace{1.3cm} -\frac{2}{3}\,\tA_{ij}\,\partial_k\beta^k+\beta^k\,\partial_k \tA_{ij}\, \label{eqn:dtAij2} ~, \\ 
&\partial_t \tilde \Gamma^i=2\,\alpha\left(\tilde\Gamma^i_{jk}\,\tA^{jk}-\frac{2}{3}\,\tilde\gamma^{ij}\partial_j K - \frac{3}{2}\,\tA^{ij}\frac{\partial_j \chi}{\chi}\right) \nonumber \\
&\hspace{1.3cm} -2\,\tA^{ij}\,\partial_j \alpha +\beta^k\partial_k \tilde\Gamma^{i} \nonumber \\
&\hspace{1.3cm} +\tilde\gamma^{jk}\partial_j\partial_k \beta^i +\frac{1}{3}\,\tilde\gamma^{ij}\partial_j \partial_k\beta^k \nonumber \\
&\hspace{1.3cm} + \frac{2}{3}\,\tilde\Gamma^i\,\partial_k \beta^k -\tilde\Gamma^k\partial_k \beta^i - 16\pi\,\alpha\,\tilde\gamma^{ij}\,S_j ~ , \label{eqn:dtgamma2}
\end{align} 
where the superscript $\rm{TF}$ denotes the trace-free parts of tensors, with $R\ij$ being the (3-dimensional) Ricci tensor.  The 3+1 decomposition of the energy-momentum tensor $T^{\mu\nu}$ gives
\bea \label{3+1sources_AP}
 \rho &=& n^\mu n^\nu T_{\mu\nu} ~,\\
 S_i &=& -\gamma^{\mu}_i n^\nu T_{\mu\nu} ~,\\ 
 S_{ij} &=& \gamma^{\mu}_i \gamma^{\nu}_j T_{\mu\nu} ~,\\ 
 S &=& \gamma\IJ S\ij ~,
\eea
where $n^\mu=(1/\alpha, -\beta^i/\alpha)$ is the unit normal vector to the three-dimensional slices.

The Hamiltonian and momentum constraints, 
\begin{align}
\mathcal{H} & = R + K^2-K_{ij}K^{ij}-16\pi \rho= 0\, , \label{eqn:HamSimp} \\
\mathcal{M}_i & = D^j (K_{ij} - \gamma_{ij} K) - 8\pi S_i =0\, , \label{eqn:MomSimp}
\end{align}
where $R$ is the Ricci scalar, are only solved explicitly when constructing initial data. However, they are also monitored during the time evolution in order to ensure that there is no significant deviations from General Relativity. 

\subsection{Gauge choice and singularity avoidance }%

The gauge parameters are initially set to $\alpha=1$ and $\beta^i=0$ and then evolved in accordance with the \textit{moving puncture gauge} \cite{Baker_2006, Campanelli_2006}, for which evolution equations are
\begin{eqnarray}
\partial_t \alpha &=& -\eta_\alpha \alpha \left( K - \langle K \rangle \right) + \,\beta^i\partial_i \alpha \ , \label{eqn:alphadriver}\\
\partial_t \beta^i &=& B^i\, \label{eqn:betadriver},\\
\partial_t B^i &=& \frac34\, \partial_t \tilde\Gamma^i - \eta_B\, B^i\ \,, \label{eqn:gammadriver}
\end{eqnarray}
where the constants $\eta_\alpha$ and $\eta_B$ are conveniently chosen to improve the numerical stability. This way, $\alpha$ and $\beta^i$ are boosted in the problematic regions with strongly growing extrinsic curvature and spatial derivatives of the three-metric $\tilde \gamma_{ij}$. 
The goal of this gauge is to prevent the code from resolving the central singularity of any black hole that may eventually form, as well as to prevent coordinate singularities on converging geodesics.

\subsection{Scalar field equations}
For the Einstein frame canonical scalar field $\varphi^I$, the energy-momentum tensor is given by
\begin{equation}
T_{\mu\nu} = \delta_{IJ}\left( \partial_\mu \varphi^I\, \partial_\nu \varphi^J - \frac{1}{2} g_{\mu\nu}\, \partial_\lambda \varphi^I \, \partial^\lambda \varphi^J \right) - g_{\mu\nu} V(\varphi^K) \,
\end{equation}

The scalar field dynamics is governed by the the Klein-Gordon equation, split into two first order equations for the field and its momentum $\Pi_{\rm M}^I$
\begin{align}
\partial_t \varphi^I &= \alpha \Pi_{\rm M} +\beta^i\partial_i \varphi \label{eq:dtvarphi} ~ , \\
\partial_t \Pi_{\rm M}^I &= \beta^i\partial_i \Pi_{\rm M}^I + \alpha\partial_i\partial^i \varphi^I + \partial_i \varphi^I \, \partial^i \alpha \\
& \ +\alpha \left( K\Pi_{\rm M}^I-\gamma^{ij}\Gamma^k_{ij}\partial_k \varphi^I - \frac{d}{d\varphi^I}V(\varphi^K) \right) ~ ,
\end{align} 

Still in the Einstein frame, but with the Jordan defined scalar fields $\bar\phi^I$, the energy momentum is written as

\beq
T_{\mu\nu} = {\cal G}_{IJ} \partial_\mu \bar\phi^I \partial_\nu \bar\phi^J - g_{\mu\nu} \left[ \frac{1}{2} {\cal G}_{IJ} \partial_\alpha \bar\phi^I \partial^\alpha \bar\phi^J + V (\bar\phi^I ) \right] ~,
\eeq

and then the evolution equations are read

\begin{align}
\partial_t \bar\phi^I &= \alpha {\bar\Pi}_{\rm M}^I +\beta^i\partial_i \bar\phi^I ~ , \\
\partial_t {\bar\Pi}_{\rm M}^I &= \beta^i\partial_i {\bar\Pi}_{\rm M}^I + \alpha\partial_i\partial^i \bar\phi^I + \partial_i \bar\phi^I \, \partial^i \alpha \\ \nonumber
& \ +\alpha \Big[ K {\bar\Pi}_{\rm M}^I- \gamma^{ij}\Gamma^k_{ij}\partial_k \bar\phi^I \\ \nonumber
& + \Gamma^I_{JK} \left( - \bar\Pi_{\rm M}^J\bar\Pi_{\rm M}^K  + \gamma\IJ\partial_i \bar\phi^J \partial_j \bar\phi^K \right)
- {\cal G}^{IJ} \frac{d }{d {\bar\phi}^J} V (\bar\phi^K) \Big] ~ .
\end{align} 

If instead, the system is evolved using the Einstein frame notation for the scalar fields $\Phi^I$, the energy tensor simplifies to 

\beq
T_{\mu\nu} = {\delta}_{IJ} \partial_\mu \Phi^I \partial_\nu \Phi^J - g_{\mu\nu} \left[ \frac{1}{2} {\delta}_{IJ} \partial_\alpha \Phi^I \partial^\alpha \Phi^J + V (\Phi^I ) \right] ~,
\eeq

and then, the evolution equations are given by

\begin{align}
\partial_t \Phi^I &= \alpha {\Pi}_{\rm M}^I +\beta^i\partial_i \Phi^I ~ , \\
\partial_t {\Pi}_{\rm M}^I &= \beta^i\partial_i {\Pi}_{\rm M}^I + \alpha\partial_i\partial^i \Phi^I + \partial_i \Phi^I \, \partial^i \alpha \\ \nonumber
& \ +\alpha \Big[ K {\Pi}_{\rm M}^I- \gamma^{ij}\Gamma^k_{ij}\partial_k \Phi^I  - \frac{d }{d {\Phi}^I} V (\Phi^K) \Big] ~ .
\end{align}

\section{Field-space metric and Christoffel symbols}
\label{sec:AppendixAFieldSpace}

Given $f ( {h} , {s} ) = \left({M^2_{\rm pl}} + \xi_ {h} {h} ^2 + \xi_ {s} {s} ^2 \right)/2$ for a two-field model, with non-minimal couplings $\xi_{h},\> \xi_{s}$, the field-space metric in the Einstein frame, takes the form
\beq
{\cal G}_{IJ}=
\left( \frac{M_{\rm pl}^2}{4f^2} \right)
\begin{pmatrix} 
2f+6\xi_ {h} ^2 {h} ^2 & 6 \xi_ {h} \xi_ {s} {h} {s} \\
6 \xi_ {h} \xi_ {s} {h} {s} & 2f+6\xi_ {s} ^2 {s} ^2
\end{pmatrix} \, ,
\label{G_hh}
\eeq
\beq
{\cal G}^{IJ}=
\left( \frac{2f}{M_{\rm pl}^2C} \right)
\begin{pmatrix} 
2f+6\xi_ {s} ^2 {s} ^2 & -6 \xi_ {h} \xi_ {s} {h} {s} \\
-6 \xi_ {h} \xi_ {s} {h} {s} & 2f+6\xi_ {h} ^2 {h} ^2
\end{pmatrix} \, ,
\label{invG}
\eeq
where $C ({h} , {s} )$ is defined as
\beq
\begin{split}
C({h} , {s}) = 2f + 6 \xi_ {h} ^2 {h} ^2 + 6 \xi_ {s} ^2 {s} ^2 .
\end{split}
\label{C}
\eeq

The Christoffel symbols for this field space take the form
\beq
\begin{split}
\Gamma^ {h}_{\>\> {h} {h} } &= \frac{\xi_{h} (1 + 6 \xi_{h} ) {h} }{C} - \frac{\xi_ {h} {h} }{f}\, , \\
\Gamma^ {h}_{\>\> {h} {s} } &= - \frac{\xi_ {s} {s} }{2f} \quad = \Gamma^ {h} _{\>\> {s} {h} } \, , \\
\Gamma^{h} _{\>\> {s} {s} } &= \frac{\xi_{h} (1 + 6 \xi_{s} ) {h} }{C} , \\
\Gamma^{s} _{\>\> {s} {s} } &= \frac{\xi_{s} (1 + 6 \xi_{s} ) {s} }{C} - \frac{\xi_ {s} {s} }{f} \, ,\\ 
\Gamma^{s} _{\>\> {s} {h} } &= - \frac{\xi_ {h} {h} }{2f} \quad = \Gamma^ {s} _{\>\> {h} {s} } \, , \\ 
\Gamma^{s}_{\>\> {h} {h} } &= \frac{\xi_{s} (1 + 6 \xi_{h} ) {s} }{C} \, . 
\end{split}
\label{Gammas}
\eeq

\section{Scalar fields in the Jordan and Einstein frame notation \label{ApConversionJE} }

Transforming from the scalar fields in the Jordan frame notation $\bar\phi^I$ to the Einstein frame once $\Phi^I$ is done by finding an approximate solution to the following system of equations

\beq \label{eq:AP_convert_frame}
{\cal G}_{IJ} g^{\mu\nu} \partial_\mu \bar\phi^I \partial_\nu \bar\phi^J = {\delta}_{IJ} g^{\mu\nu} \partial_\mu \Phi^I \partial_\nu \Phi^J
\eeq

Assuming two Jordan scalar fields, the Higgs $h$ with non-minimal coupling $\xi_h$  and an auxiliary field $s$ with non-minimal coupling $\xi_s$. Then we search for a transformation into the Einstein frame such as $\varphi(h, s),\> \chi(h,s)$. Assuming $\xi_s = 0$ , the above mentioned system of equation simplifies to
\begin{align}
 \left( \frac{\partial \varphi}{\partial h} \right)^2 &+ \left( \frac{\partial \chi}{\partial h} \right)^2 = {\cal G}_{hh} \\  
  \left( \frac{\partial \varphi}{\partial s} \right)^2 &+ \left( \frac{\partial \chi}{\partial s} \right)^2 = {\cal G}_{ss} \\ 
  \left( \frac{\partial \varphi}{\partial h} \frac{\partial \varphi}{\partial s} \right) &+ \left( \frac{\partial \chi}{\partial h} \frac{\partial \chi}{\partial s} \right) = 0 ~. 
\end{align}

\begin{figure}[b]
\begin{center}
\hspace*{-2mm}
\includegraphics[width=8.5cm]{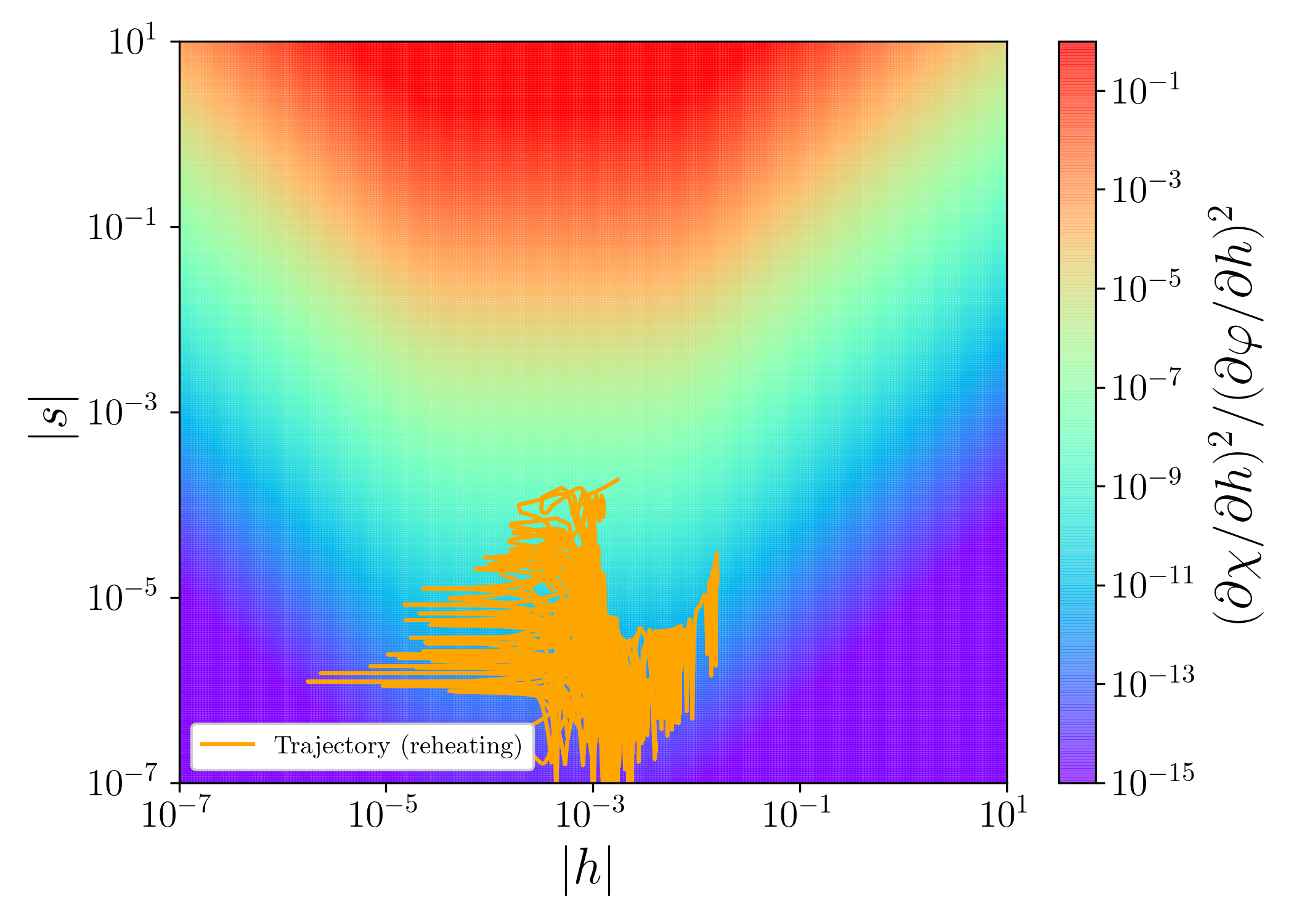}
\end{center}
\caption{ Error of the approximation (\ref{eq:AP_convert_frame}) to convert fields between the Jordan and Einstein frame.  \label{fig:approx}}
\end{figure}

By assuming $ {\partial \varphi}/{\partial h} \approx  \sqrt{{\cal G}_{hh}}$ and $ {\partial \chi}/{\partial s} \approx  \sqrt{{\cal G}_{ss}}$ implies that, in the range of validity of this approximation, 
\begin{align}
\left(\frac{\partial \varphi}{\partial s}\right) ^2 &\ll \left(\frac{\partial \chi}{\partial s}\right)^2  \label{AP_cond1}
\\ 
\left(\frac{\partial \chi}{\partial h}\right)^2    &\ll
\left(\frac{\partial \varphi}{\partial h}\right)^2 
\label{AP_cond2} ~.
\end{align}
While the first identity is trivially satisfied, the second one  (\ref{AP_cond2}) is not, because the approximation proposes the solution to be {$s \approx \chi \sqrt{2 f(h)}$}. Thus, the validity of these approximation  depend on the region in consideration of the field space, as shown in shown in Fig. \ref{fig:approx}. In general, the parameter space when this assumptions are valid is generally at when  $ |s| <10^{-2}$, and when  $ |s| < |h| / 100$.  These regions corresponds to $ 
\left(\frac{\partial \chi}{\partial h}\right)^2 / \left(\frac{\partial \varphi}{\partial h}\right)^2  < 10^{-5}$.

\section{Code validation and convergence tests}

The validation of the code is done by monitoring the constraint equations.  The relative Hamiltonian constraint is defined as follows
\be
  \mathcal{H}_{\rm REL} = \frac {\mathcal{H} }{\left[\mathcal{H}\right] }  
\ee
where 
\be
\left[\mathcal{H}\right]  \equiv \Biggl[  \lp {R} \rp ^2 +  \lp \tilde A\IJ \tilde A\ij \rp ^2 
+ \lp  \frac23 K^2 \rp ^2  + \lp 16\pi \rho_{\rm sf}\rp ^2 \Biggr] ^{1/2}
\ee

\begin{figure*}[h]
\begin{center}
\includegraphics[width=7.5cm]{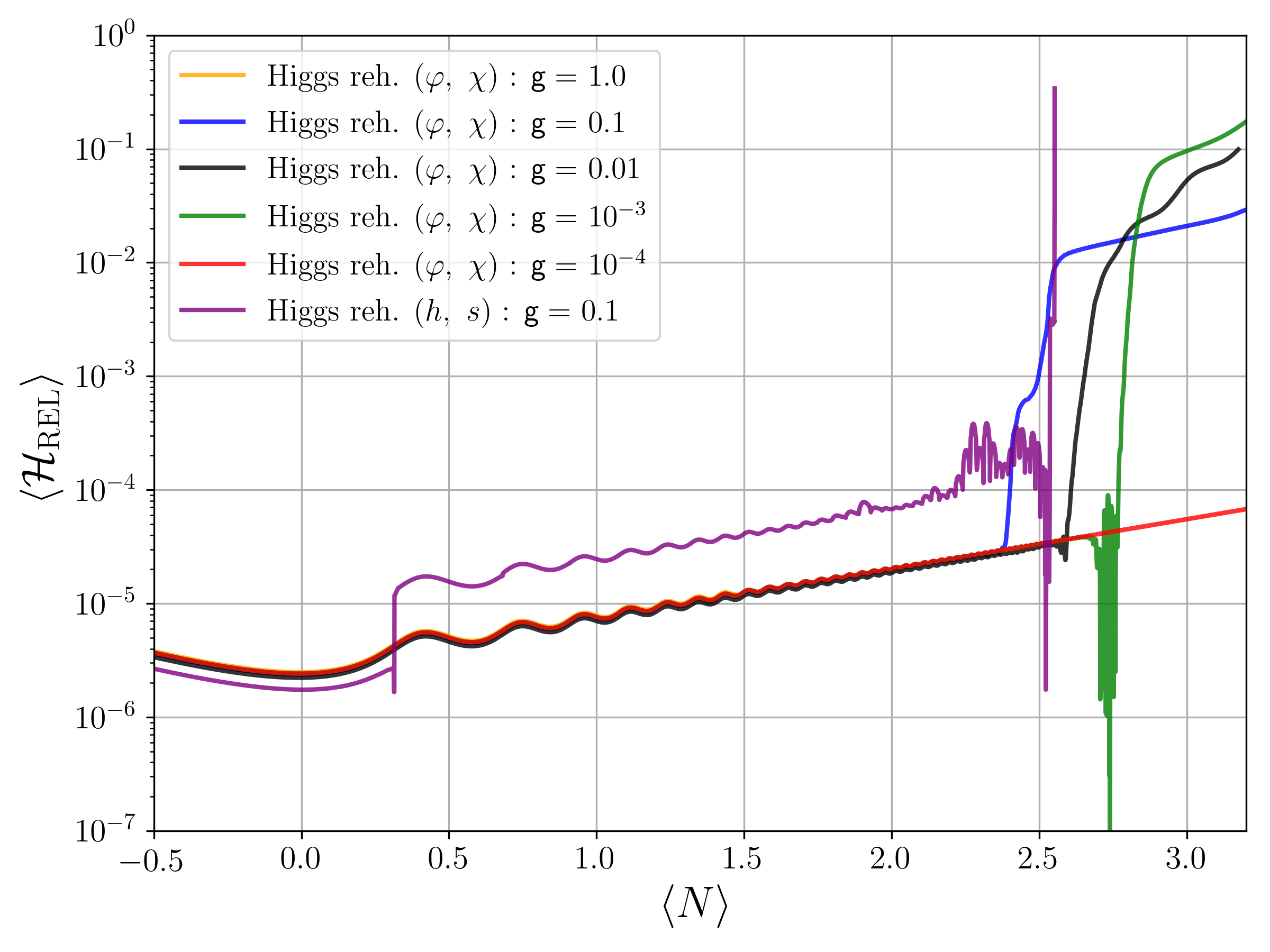}
\includegraphics[width=7.5cm]{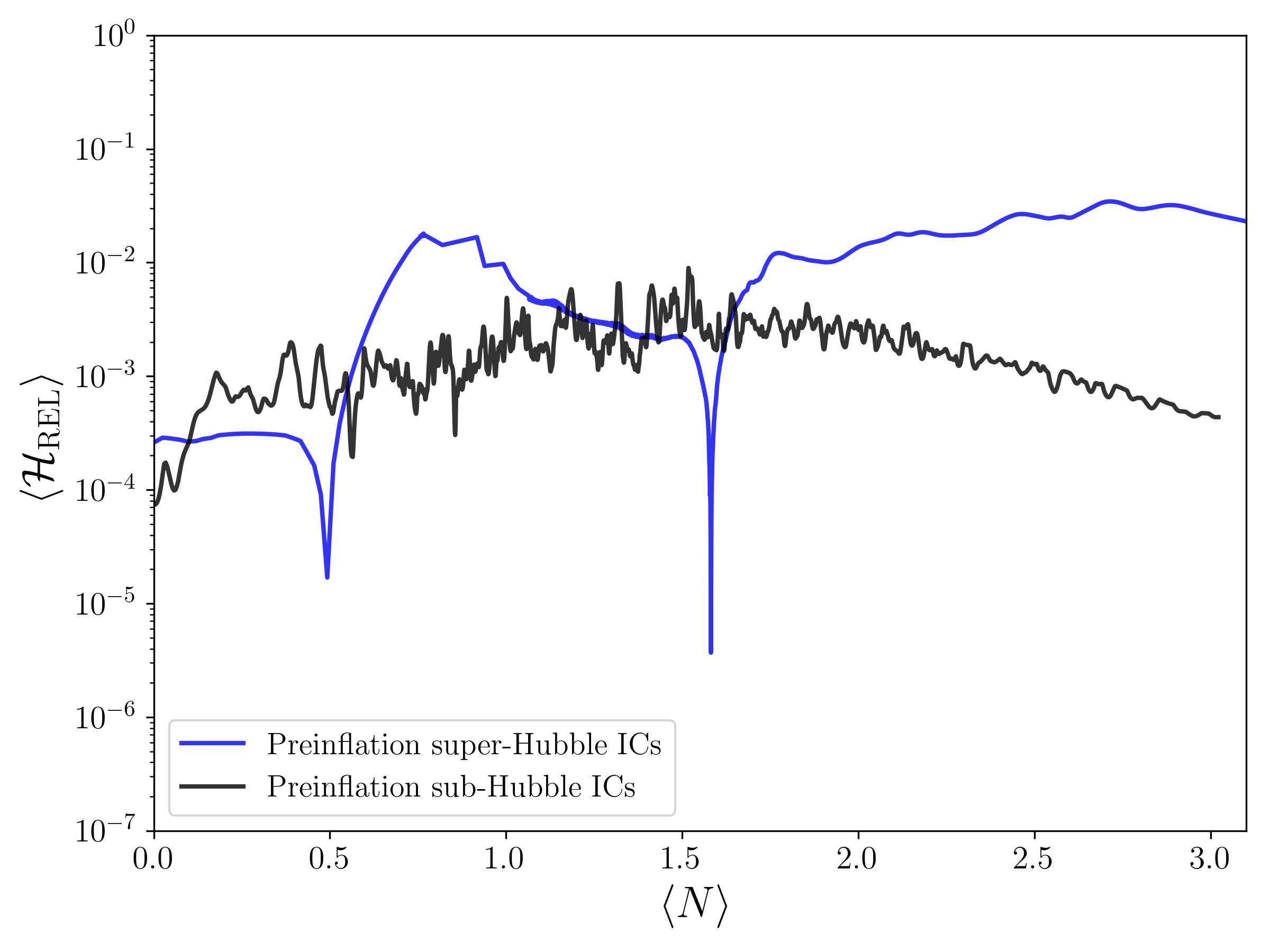}
\end{center}
\caption{ Relative Hamiltonian constraint for simulations on reheating (left panel) and on preinflation (right panel).  \label{fig:HamVal} }
\end{figure*}

\begin{figure*}[t]
\begin{center}
\includegraphics[width=7.5cm]{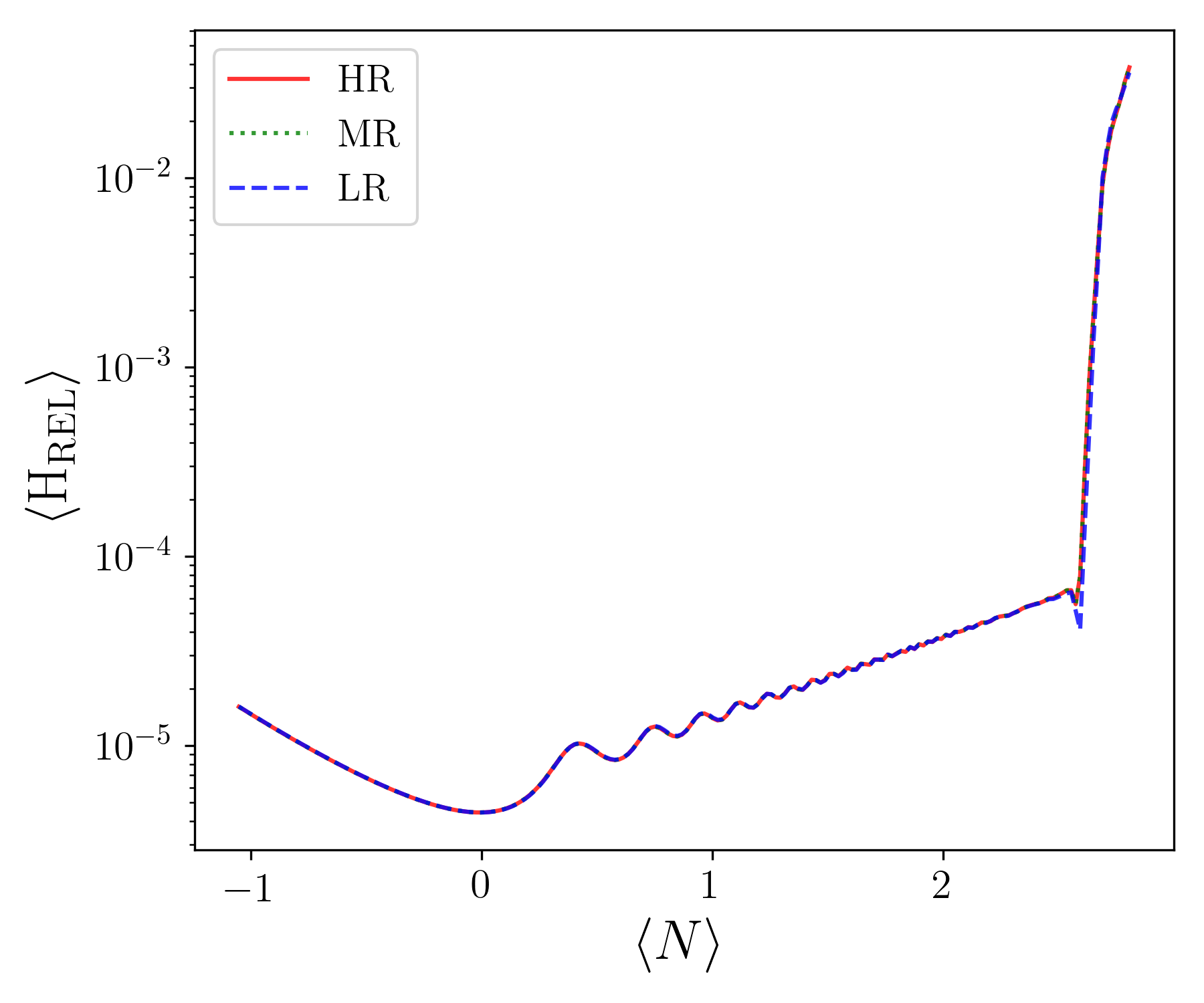}
\includegraphics[width=7.5cm]{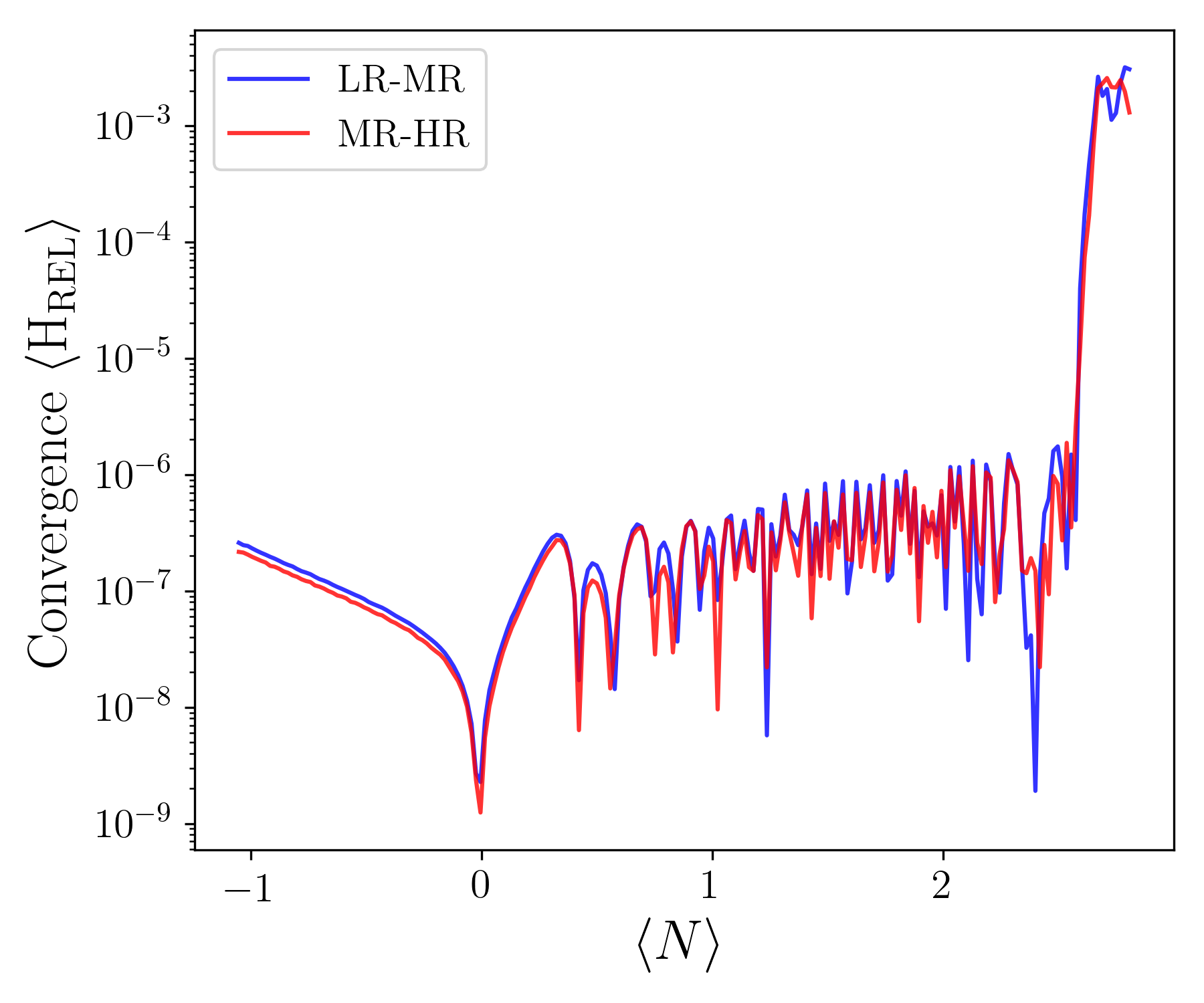}
\includegraphics[width=7.5cm]{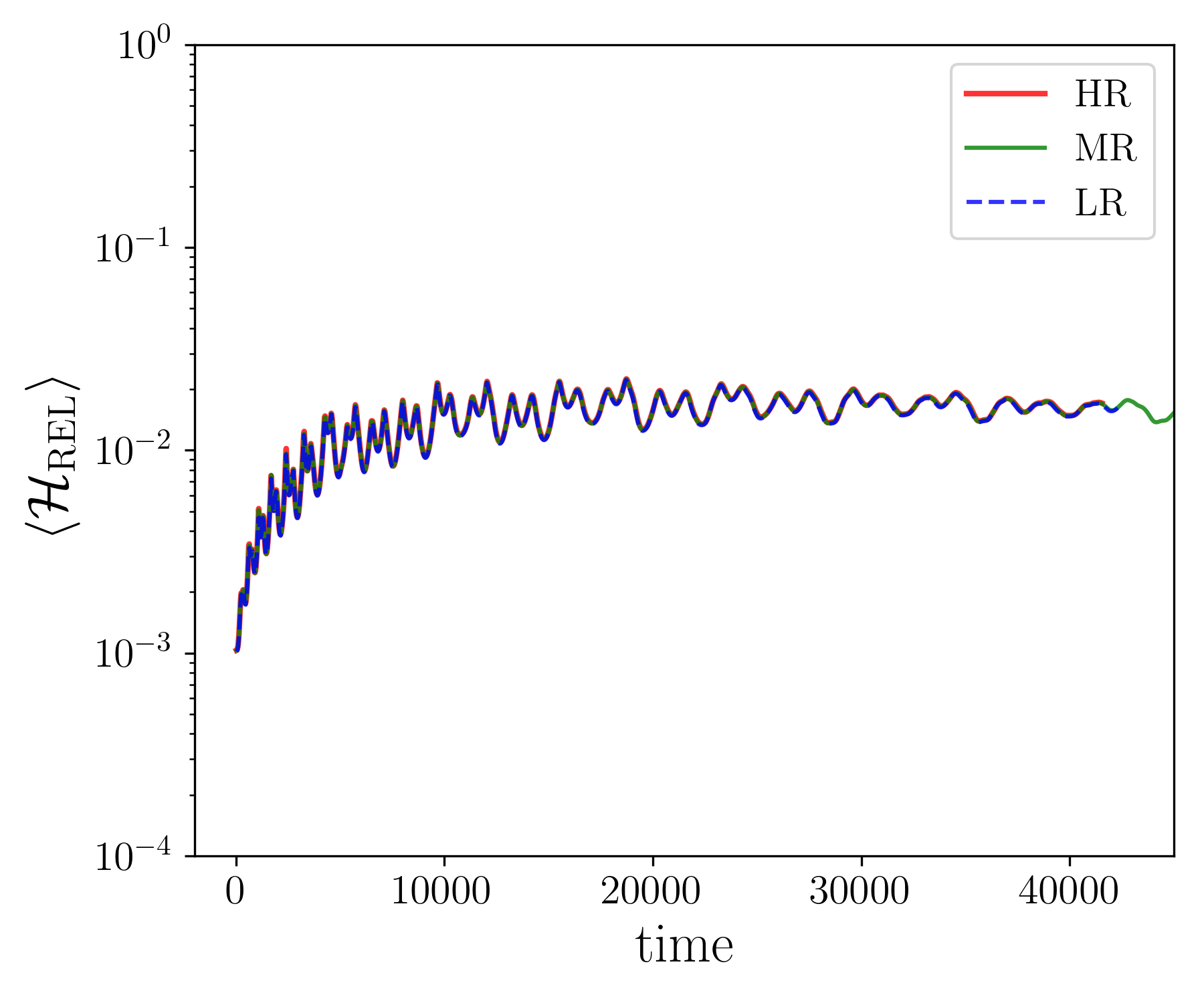}
\includegraphics[width=7.5cm]{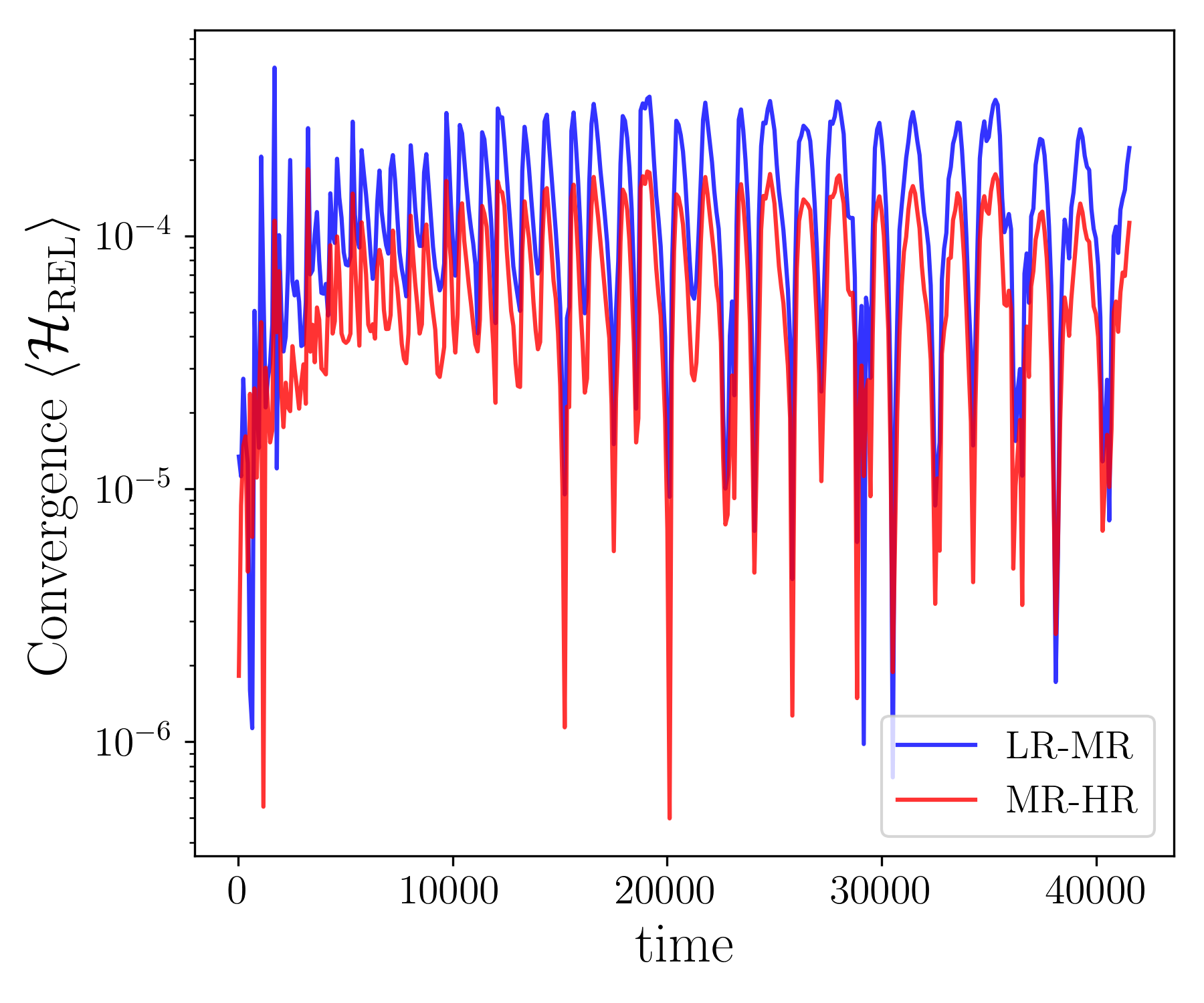}
\end{center}
\caption{ Convergence testing: Relative difference  between low (LR), medium (MR) and high (HR) resolutions grids for simulations on preheating (top panels) and preinflation (bottom panels).  The size grid used is LR = $128^3$, MR = $144^3$, HR = $156^3$ for the case of preheating, and   LR = $128^3$, MR = $180^3$, HR = $220^3$ for the case of preinflation. 
\label{fig:ctest} }
\end{figure*}

These quantity has been computed for all simulations, which is shown in Fig. \ref{fig:HamVal}. Convergence tests using different grid-size resolutions are shown in Fig. \ref{fig:ctest}.

\ \\

\section{Auxiliary figures \label{App:AuxiliarlyFigs}}

\begin{figure*}[t!]
\begin{center}
\includegraphics[width=15.5cm]{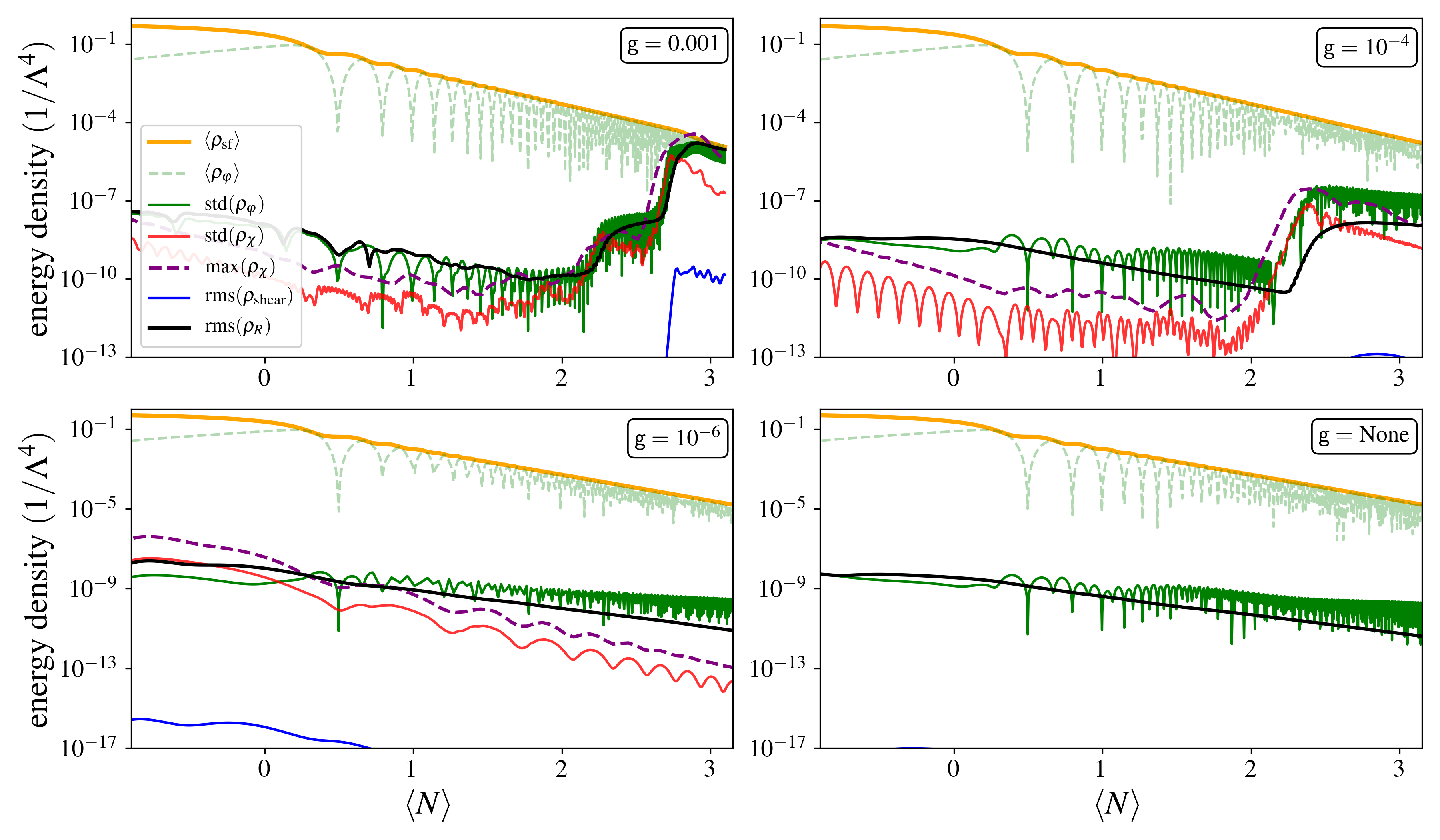}
\end{center}
\caption{ Evolution of the energy density respect on the expansion history. This simulations corresponds to scenarios of preheating with $\mathsf{g} \leq 10^{-3}$ from  Fig. \ref{fig:reheating2}, but including perturbations in the initial state of the Higgs field. The bottom-right panel corresponds to the single-field case.  The box size of the simulations at the end of inflation correspond to $L\approx 5 H^{-1}$.
\label{fig:aux_reheating}
}
\end{figure*}

\begin{figure*}[t!]
\begin{center}
\includegraphics[width=15.5cm]{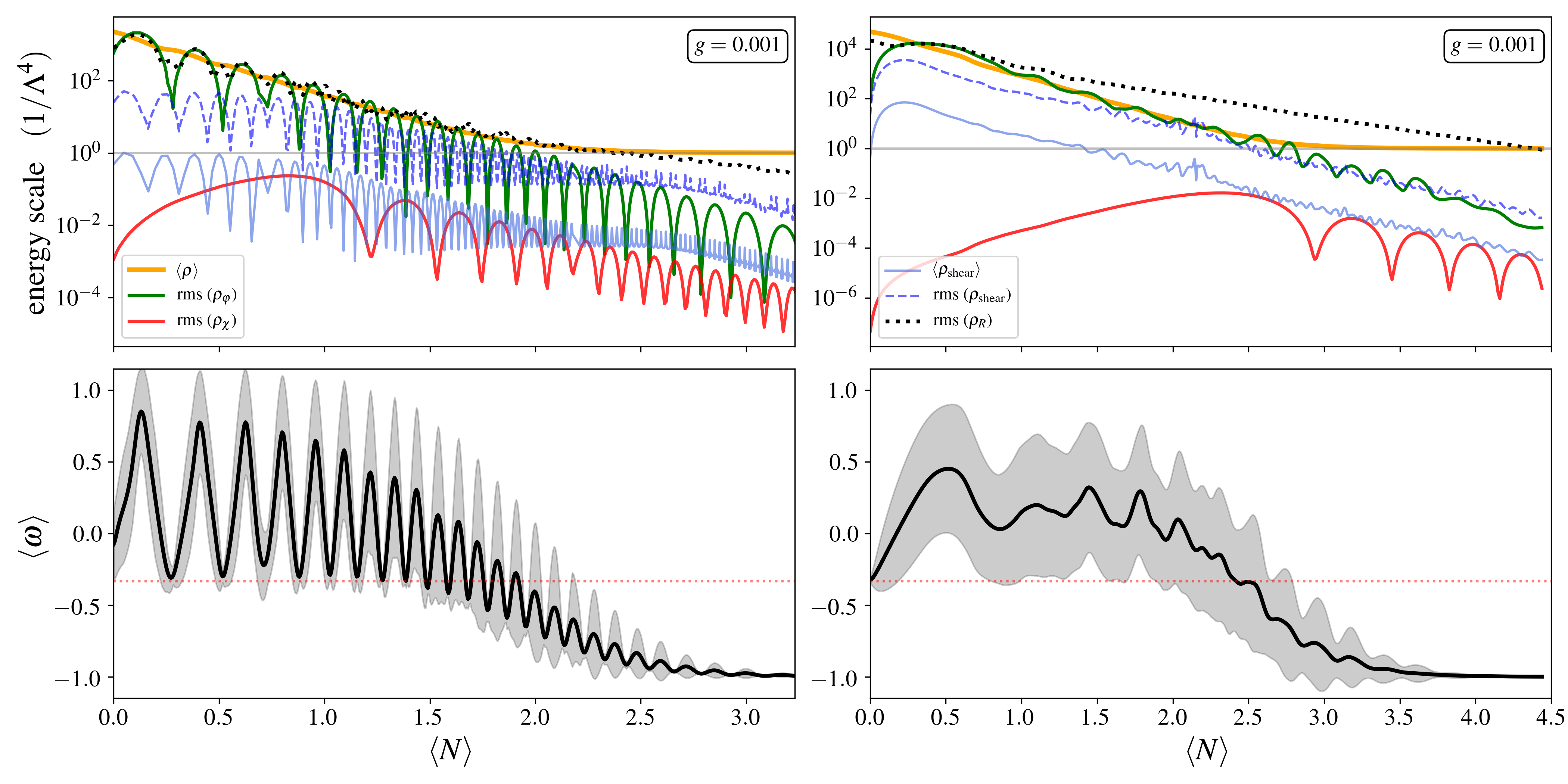}
\end{center}
\caption{ 
Same as in top and bottom panels of Fig. \ref{fig:preinflation}. It shows the dynamical evolution of sub-Hubble (left) and super-Hubble (right) energetically dominated initial conditions corresponding to the pre-inflationary era until the onset of inflation. The initial box size of the simulations correspond to $L\approx 2 H^{-1}$ for the sub-Hubble case and $L\approx 10 H^{-1}$ for the super-Hubble one.
\label{fig:aux_preinf}
}
\end{figure*}

In this appendix, we include additional figures corresponding to alternative simulations constructed with some variation in the initial conditions. For the simulations on preheating, we tested whether the inclusion of initial perturbations in the Higgs field induce changes during the resonance period, in particular in the low $\mathsf{g}$ limit as perturbations in the Higgs field during the last efold of inflation could have been diminished. In addition, a larger initial box size was also considered. These modifications, as shown in Fig.~\ref{fig:aux_reheating}, do not significantly change the resonance dynamics of the preheating process.  Similarly, alternative initial conditions for the pre-inflationary era are shown in Fig.~\ref{fig:aux_preinf}, where larger box sizes where tested, as well as different patterns in the initial field gradients. These simulations also lead to the same conclusions explained in  section~\ref{sec:OnInitialConditions}.

\clearpage

\end{document}